\documentclass[12pt]{article}

\usepackage{amsmath,amsthm,amssymb}

\usepackage{feynmp-auto}
\usepackage{xcolor}
\usepackage{upgreek}

\usepackage{graphicx}
\usepackage{dcolumn}
\usepackage{bm}
\usepackage{amsmath} 

\usepackage{caption} 
\usepackage{subcaption}

\begin{document}
\begin{fmffile}{diagram}

\begin{center}
{\bf \Large  Lepton universality in a model with three generations of sterile Majorana neutrinos}
\end{center}

\begin{center}
{M.N. Dubinin$^{a,b}$, D.M. Kazarkin$^{c}$\footnote{corresponding author, e-mail: {\tt kazarkin.dm17@physics.msu.ru}} \\
\hfill\\
{\small \it $^a$Skobeltsyn Institute of Nuclear Physics, Lomonosov Moscow State University,}\\
{\small \it 119991, Moscow, Russia} \\
{\small \it $^b$National University of Science and Technology MISIS,\\
119049, Moscow, Russia}\\
{\small \it $^c$Physics Department, Lomonosov Moscow State University}\\
{\small \it 119991, Moscow, Russia}
}
\end{center}

\begin{center}
{\bf Abstract}
\end{center}
\begin{quote}

The extension of the Standard Model lepton sector by three right-handed Majorana neutrinos (heavy neutral leptons, HNL) with masses up to GeV scale is considered. While the lightest heavy neutral lepton is the dark matter particle with mass of the order of 5 keV, the remaining two HNLs ensure standard (active) neutrino mass generation by means of the see-saw type I mechanism. Two heavy sterile neutrinos with quasi-degenerate masses up to 5 GeV can induce the deviation of lepton universality violation parameter in the decays of $\pi^\pm$ and $K^\pm$ mesons from the the Standard Model value.
Contours are obtained for the permissible values of this parameter within the framework of two mixing scenarios, taking into account the lifetime boundary for heavy neutral lepton from Big Bang nucleosynthesis in the Universe. When calculating the HNL decay width in the framework of the model with six Majorana neutrinos, three active and three heavy, both two-particle and three-particle lepton decays, essential for masses below the mass of the pion, were taken into account. When calculating the decay widths, the limiting case known as the "Dirac limit" is not used. The results based on the explicit form of mixing matrices for three HNL generations and the diagram technique for Majorana neutrinos, which explicitly take into account the interference terms for diagrams with identical mass states, can lead to some differences in lifetime from the results using the "Dirac limit" and the displacement of the corresponding experimental exclusion contours of the "mass-mixing" type. For the second mixing scenario, a mass region of $460~\text{MeV} <M< 485$ MeV has been found that allows violation of lepton universality in charged kaon decays at the level observed in the experiment.
\end{quote}   

\newpage

\section{Introduction. The main features of the model.}

Extension of the Standard Model (SM) leptonic sector by heavy neutral leptons of right helicity (HNL, also referred to as sterile Majorana neutrinos) known for a long time \cite{sterile1, sterile2} has been analyzed multilaterally recently due to the attractive general features of such an extension, within the framework of which the symmetry between right and left neutrinos is restored, new large energy scales are not necessarily introduced, 
neutrino oscillations and their masses generation by means of the see-saw mechanism are successfully explained and a number of important cosmological applications of the model, such as baryon asymmetry of the Universe generation, description of the inflationary stage of the early Universe and its accelerated expansion at the present time are successfully interpreted \cite{general1,hnl_dm}. 

New useful features are realized in the construction of the so-called minimal neutrino standard model $\nu$MSM \cite{shaposhnikov1,shaposhnikov2} which is a minimal extension of the SM. In $\nu$MSM framework the HNL masses do not exceed the electroweak scale and there are no other new particles up to the Planck scale. Cosmological observations impose significant limitations on the model parameter space, which lead to at least three HNL and establish a strict upper limit on the mass of the lightest active neutrino $min(\nu_{e,\mu,\tau})$. The lightest heavy lepton $N_1$ with mass of the order of 10 keV, the lifetime more than $\tau_{\rm Universe} \sim 10^{17}$ sec and mixing parameter of the order of 10$^{-13}$ -- 10$^{-7}$, plays a role of the dark matter (DM) particle in such an extension \cite{boyarski, Merle:2013gea}. The direct method of $N_1$ DM detection is due to the possibility of observation of the one-loop decay process of $N_1 \to \gamma \nu$ in galactic media \cite{nugamma}. Two remaining heavy leptons ensure the mechanism of mass generation of standard (or active) neutrinos, their masses can vary in a wide range of values up to multiGeV scale, however, enough baryon asymmetry through oscillation-induced leptogenesis can be generated even if they are of the order of MeV and their mass splitting is rather small \cite{asaka3}.  

Mixing of light enough HNL states with active neutrino states could lead to observable HNL production in charged meson decays such as $\pi^+ \to e^+ N_{2,3}$ and $K^+ \to l^+, N_{2,3}$, $l=e,\mu$ which could violate lepton universality principle demonstrating departures from the SM ratio $R_{\pi,K}=\Gamma(\pi,K\to e\nu)/\Gamma(\pi,K \to \mu \nu)$ \cite{shrock1, shrock2, abada1, abada2, asaka} which is a quantity stable with respect to radiative corrections and hadronization uncertainties.  

In this paper we estimate the possible departure of the lepton universality parameter 
    \begin{equation} \label{luv_param}
        \Delta r_M = \frac{R_M}{R^{SM}_M}-1
    \end{equation}
from zero value due to HNL contributions in the $\nu$MSM-like model where an explicit form of mixing for the three lepton generations is used. The Lagrangian of extension has the form
    \begin{equation}
        \mathcal{L} = \mathcal{L}_{SM} + i \overline{\nu}_R \partial_\mu \gamma^\mu \nu_R- \left( F \; \overline{l}_L \nu_R \tilde{H} + \frac{M_{M}}{2} \overline{\nu^c}_R \nu_R + h.c \right),
        \label{lagr}
    \end{equation}
where
$l_L=(\nu_L, \, e_L)^T$ is the left lepton doublet, 
$\nu_R$ are HNL flavor states, $(\nu_R)^c \equiv C \overline{\nu}_R^T$ ($C=i \gamma_2 \gamma_0$), $\overline{\nu}_R \equiv \nu_R^\dagger \gamma^0$,  
$H$ is the Higgs doublet ($\tilde{H}=i\tau_2 H^\dagger$),
$F$ is the Yukawa matrix and $M_{M}$ is a Majorana mass matrix. After spontaneous symmetry breaking $M_{D}= F \langle H \rangle=F v$ ($v=174$ GeV) is the matrix of Yukawa term. The full 6$\times$6 mass matrix
\begin{eqnarray}
    \frac{1}{2}(\overline{\nu}_L \vspace{2mm} \overline{\nu^c}_R)
        \mathcal{M}
    \left(
        \begin{array}{c}
        \nu_L^c \\
        \nu_R
        \end{array}
        \right)    +h.c.
        \, = \, \frac{1}{2} (\overline{\nu}_L \vspace{2mm} \overline{\nu^c}_R) \left(
        \begin{array}{cc}
         0 & M_D\\
         M_D^T & M_M  
        \end{array}
        \right)
        \left(
        \begin{array}{c}
        \nu_L^c \\
        \nu_R
        \end{array}
        \right) +h.c.,
        \label{M66}
\end{eqnarray}
where the flavor states $(\nu_{L})_\alpha, (\nu_{R})_I$ and the mass states $\nu_k, N_I$ ($\alpha = e,\mu,\tau$, $k,I=1,2,3$) are connected by the unitary transformation
\begin{eqnarray}
    \left(
    \begin{array}{c}
    \nu_L \\
    \nu_R^c 
    \end{array} \right)
    =\mathcal{U} P_L \left(
    \begin{array}{c} 
    \upnu \\
    N 
    \end{array}
    \right),
    \hskip 8mm \mathcal{U}= \mathcal{W} \cdot {\rm diag}(U_\nu, U_N^*) 
\end{eqnarray}
where $P_L$ is the left projector and $U_\nu, U_N$ are unitary $3 \times 3$ matrices. Block-diagonal form of the mass matrix \eqref{M66} looks as
\begin{eqnarray}
    \mathcal{U}^\dagger \mathcal{M} \mathcal{U}^*
    =
    \left(
    \begin{array}{cc}
     U_\nu^\dagger & 0 \\
     0 & U_N^T  
    \end{array}
    \right) 
    \mathcal{W}^\dagger \mathcal{M} \mathcal{W}^*
    \left(
    \begin{array}{cc}
     U_\nu^* & 0 \\
     0 & U_N  
    \end{array}
    \right)
    =\left(
    \begin{array}{cc}
     U_\nu^\dagger m_\nu U_\nu^* & 0 \\
     0 & U_N^T M_N U_N  
    \end{array}
    \right)
    =
    \left(
    \begin{array}{cc}
     \hat{m} & 0 \\
     0 & \hat{M}  
    \end{array}
    \right),
\end{eqnarray}
where
$\hat{m} = {\rm diag}(m_1, m_2, m_3)$, $\hat{M} = {\rm diag}(M_1, M_2, M_3)$, $\mathcal{W}^\dagger \mathcal{M} \mathcal{W} = diag(m_\nu, \; M_N)$, $\mathcal{M}$ is defined by Eq.\eqref{M66}.
In the following diagonalization procedure \cite{ibarra1} the unitary $\mathcal{W}$-matrix is represented as an
exponent of an antihermitian matrix 
\begin{equation} 
\label{ow}
    \mathcal{W} = \exp \left(
    \begin{array}{cc}
        0 & \theta \\
        -\theta^\dagger & 0
    \end{array}\right)    
\end{equation}
and decomposed
\begin{eqnarray}
        \mathcal{W} = 
            \left(
            \begin{array}{cc}
            1-\frac{1}{2} \theta \theta^\dagger + O(\theta^4) & \theta + O(\theta^3)\\
            -\theta^\dagger + O(\theta^3) & 1-\frac{1}{2} \theta^\dagger \theta+O(\theta^4) 
            \end{array}
        \right).
     \label{omega}
\end{eqnarray}
The flavor states are related to the mass states in the following form
    \begin{eqnarray}
        \nu_L & \simeq & \left( 1- \frac{1}{2} \theta \theta^\dagger \right) U_\nu P_L \upnu+ \theta U_N^* P_L N, 
        \label{nuL-3} 
        \\
        \nu^c_{R} & \simeq & -\theta^\dagger U_\nu P_L \upnu + \left(1-\frac{1}{2} \theta^\dagger \theta \right) U_N^* P_L N.
    \end{eqnarray}
For the left neutrino the main contribution in the flavor basis is given by the first term in \eqref{nuL-3} which corresponds to the the well-known phenomenological relation $\nu_{L \alpha} = \sum_{\alpha} (U_{\rm PMNS})_{\alpha j} P_L \upnu_{j}$ \cite{pmns}. Deviation from unitarity for the PMNS matrix in the approximations $\mathcal{W} \sim \mathcal{O}(\theta^2)$ (and also $\mathcal{W} \sim \mathcal{O}(\theta^3)$) is given by $U_{\rm PMNS} \simeq \left( 1- \frac{1}{2} \theta \theta^\dagger \right) U_\nu$ and
defined by the $\eta=-\frac{1}{2}\theta \theta^\dagger$-matrix. The lagrangian terms for HNL currents interaction with $W^\pm,Z$ bosons have the form
    \begin{eqnarray} \label{ncurr}
        \mathcal{L}_{NC}^\nu &=& - \frac{g}{2 c_W} \gamma^\mu \overline{\upnu}_L U_{\rm PMNS}^\dagger U_{\rm PMNS} \upnu_L Z_\mu,\\  \nonumber
        \mathcal{L}_{CC}^\nu &=& -\frac{g}{\sqrt{2}}\overline{l}_L \gamma^\mu  U_{\rm PMNS} \upnu_L W_\mu^- + h.c.,\\ \nonumber
        \mathcal{L}_{NC}^N &=& -\frac{g}{2 c_W} \overline{N}_L \gamma^\mu U_N^T \theta^\dagger \theta U_N^* N_L Z_\mu\\ \nonumber
        &-& \left[ \frac{g}{2 c_W} \overline{\upnu}_L \gamma^\mu U_\nu^\dagger \theta U_N^* N_L Z_\mu +h.c. \right],\\ \nonumber 
        \mathcal{L}_{CC}^N &=& -\frac{g}{\sqrt{2}} \overline{l}_L \gamma^\mu \theta U_N^* N_L W_\mu^-+ h.c.
    \end{eqnarray}
In the following consideration we are keeping only the first and the second order terms in $\theta$, as it is customary to do in the available literature. 
The HNL mixing parameter is defined in the approximation $\mathcal{W} \sim \mathcal{O}(\theta^2)$ as  
$\Theta  \equiv  \theta U_N^*$. The standard set of active neutrino masses is defined in the framework of the $\mathcal{O}(\theta^2)$ scenario as a solution of see-saw type I equation
    \begin{equation}
        m_\nu  \simeq  - M_D \theta^T \simeq  - M_D M^{-1}_M M_D^T
    \end{equation}
with ambiguous definition of $M_D$ by means of the $U_N$ mass matrix in the HNL sector \cite{ibarra1, ibarra2}
    \begin{equation}
        \label{eq:md-omega-un}
        M_D = \pm i U_{\rm PMNS} \sqrt{\hat{m}} \Omega \sqrt{\hat{M}} U_N^\dagger,
    \end{equation}
where $\Omega$ is an arbitrary orthogonal matrix, $\Omega \Omega^T=I$.

In conclusion of this Section, simplifying for greater clarity the model to one generation of neutrino $\nu$, we recall the terminology used in the literature (see \cite{hnl_dm}) for the limiting cases of the mass term parameters in the Lagrangian, Eq.(\ref{lagr}). Case $M_M=$0 is called the "Dirac limit", since the Weyl spinors $\nu_L$ and $\nu_R$ represent the left- and right-chiral components of the Dirac neutrino with the mass term $M_D \bar \nu \nu = M_D (\bar \nu_L \nu_R+\bar \nu_R \nu_L)$. The lepton number is preserved. Case $M_M << M_D$ is called the pseudo-Dirac limit, since it is possible to divide the four components of the Dirac neutrino into two Majorana neutrinos with left-chiral components $(\nu_L \pm \nu^c_R)/\sqrt{2}$. Case $M_M \geq M_D$ corresponds to the case of seesaw mechanism ($M_M >> M_D$ is the "seesaw limit"), since we have two Majorana mass states, one of which has a mass $m_1$ of the order of $M^2_D/M_M$, the other $m_2$ of the order of $M_M$, and the mixing parameter of the two flavor states is of the order of $M_D/M_M$. To simplify calculations, the Dirac limit is used in the available literature when Feynman rules for processes involving active neutrinos and HNL are an analogue of Standard Model rules. In this paper, the Dirac limit is not used, the corresponding diagram technique for processes involving Majorana fermions was developed in \cite{denner} and \cite{haber} for calculations within supersymmetric models and can be directly applied to HNL production and decays.

For further analysis of the lepton universality within the framework of two characteristic mixing scenarios (Section 3) compatible with cosmoligical limitations (Section 2), calculations are reproduced for two-particle semileptonic HNL decays (Section 4) and calculations are made for three-particle leptonic HNL decays (Section 5) in the model with all six Majorana neutrinos. They are used for the lifetime restrictions on the mixing parameter space (Section 6). Bounds on the characteristic lepton universality parameter, Eq.\eqref{luv_param}, in the decays of $K^\pm$ and $\pi^\pm$ are considered in the framework of characteristic mixing scenarios in Section 7. 


\section{The lightest  HNL as a candidate for the role of a Dark Matter particle}

In the following it is assumed that heavy neutral leptons $N_{1,2,3}$ are ordered by mass and $N_1$ is the lightest one. For a mass $M_1$ of the order of keV the main decay channel is $N_1 \to \upnu \upnu \upnu$. The decay width corresponding to the four-fermion effective Lagrangian defined by Eq.\eqref{ncurr} has the form
    \begin{equation}
        \label{eq:width1}
    	\Gamma\Big(N_1\rightarrow \upnu \upnu \upnu \Big) = \frac{G_F^2M_{1}^5}{96\pi^3}\sum_\alpha |\Theta_{\alpha1}|^2,
    \end{equation}
where $\alpha = e,\mu,\tau$. Details of calculation can be found in the Appendix A. 

\begin{figure}[!t]
    		\includegraphics[scale=0.7]{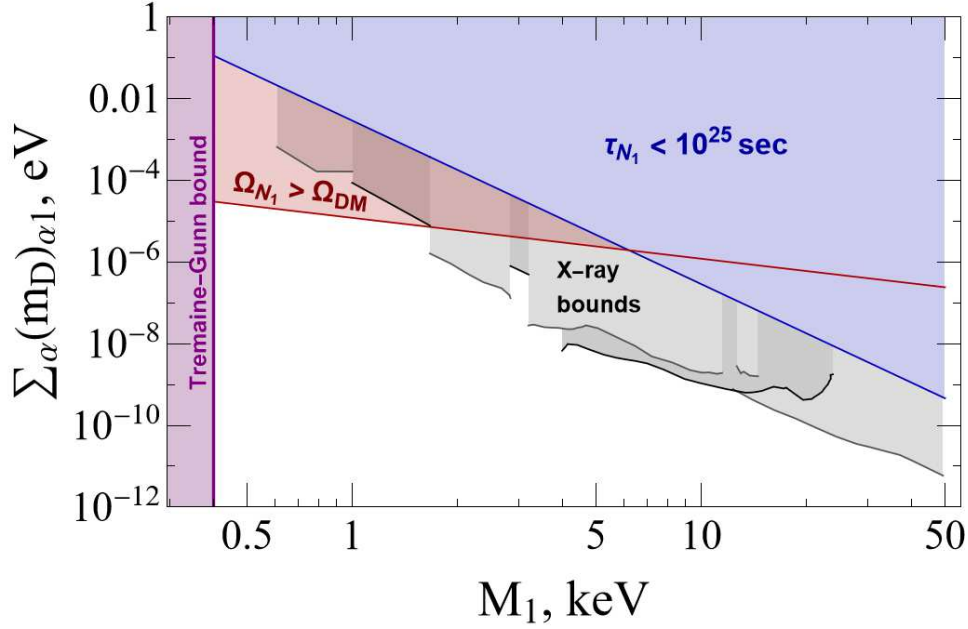}
    		\caption{Bounds for the effective mass parameter $\Sigma_\alpha (m_D)_{\alpha 1}$, see \eqref{eq:md:def}, summed by the flavour index, as a function of dark matter particle mass. See Eq.\eqref{finetuning} for the fine-tuning of mixing. Blue region is  excluded at the level $\tau_{N_1} > 10^{25}$ seconds by gamma-astronomical non-observation of $N_1 \to \gamma \upnu$ decay. Grey areas are more detailed exclusions from HEAO-1, XMM and Chandra experiments which are recalculated for $m_D$ using the contours in \cite{gamma_astro}. Violet area is excluded by Tremaine-Gunn bound \cite{Tremaine:1979we}. There is an overproduction of dark matter by means of Dodelson-Widrow mechanism \cite{DW} in the excluded light-red area above the red line.}
    		\label{fig:1}
    \end{figure}

Heavy lepton $N_1$ must not decay at time scale of order of the age of the Universe, which means $\tau_{N_1} \ge 4\times10^{17} \mbox{~sec}$. This limitation is significantly strengthened when taking into account the one-loop induced decay $N\rightarrow\gamma,\nu$, which can give a distinctive signal with photon energy $E_\gamma=M_1/2$.
The decay width
    \begin{equation} \label{eq:width2}
    	\Gamma\Big(N_1 \rightarrow \gamma,\nu\Big) = \frac{9 \alpha_{EM} G_F^2 M_1^5}{256 \pi^4} \sum_\alpha |\Theta_{\alpha1}|^2.
    \end{equation}
Although the increase of the width is small $\Gamma_{N\rightarrow \nu\nu\nu}/\Gamma_{N \rightarrow \gamma\nu}\equiv k_{rad}= \frac{8 \pi}{27 \alpha_{EM}} \approx 128$, the limitation on the lifetime can be increased by the eight orders of magnitude due to specifics of the gamma-astronomical observations, see \cite{vysotskii,xray}, providing $\tau_{N_1} > 10^{25}$ seconds.
It is convenient to introduce the {\it effective mass parameter}
    \begin{equation}
    \label{eq:md:def}
        (m_D)_{\alpha I} = \bigg|\sum_{k}\sqrt{m_k}~U_{\alpha k}\Omega_{k I}\bigg|^2
    \end{equation}
allowing to associate the masses of active neutrinos with the mixing matrix $\Theta$. The connection of the effective mass parameter with the phenomenological value of mixing $U_I^2$ is given by
    \begin{equation}
        \label{eq:md:pheno}
        \sum\limits_\alpha (m_D)_{\alpha I} = M_I U_I^2,~~\text{where}~~ U_I^2=\sum\limits_\alpha |\Theta_{\alpha I}|^2.
    \end{equation}
    Then the $N_1$ lifetime in seconds can be expressed as
        \begin{equation}
        \label{eq:lifetime_sec}
    	\tau_{N_1} = 3 \times 10^{22}\left(\frac{M_1}{1\text{~keV}}\right)^{-4} \left(\frac{\sum \limits_{\alpha}(m_D)_{\alpha 1}}{1\text{~eV}}\right)^{-1} \text{~sec},
    \end{equation}
and the gamma-astronomical constraint can be rewritten as
    \begin{equation}
    \label{eq:md:xray}
        \overline{(m_D)}_{\text{X-ray}} \equiv 3\times 10^{-3} \left(\frac{M_1}{\text{1~keV}}\right)^{-4} \text{~eV}
    \end{equation}
 where we used an estimate for the lifetime $\tau_X = 10^{25}$ seconds. It is shown by solid blue line in Fig.\ref{fig:1}.

A known direct constraint from below on the HNL mass is $M_1 > 0.4$ keV, since the distribution of HNL as fermionic dark matter in the phase space of the galaxy is limited by the distribution for a degenerate Fermi gas (Tremaine-Gunn bound, see \cite{Tremaine:1979we}).

The cosmological restriction in for the density of $N_1$ dark matter in the Universe appears in the scenario where the mixing of active and sterile neutrinos $\Theta$ is quite small, and the sterile neutrino has never been in thermal equilibrium. The dominant mechanism of the formation of sterile neutrinos (Dodelson-Widrow mechanism, see \cite{DW} ) arises from the active -- sterile neutrino oscillations. The energy fraction of sterile neutrinos in the Universe in the case of non-resonant production \cite{Abazajian:2001nj, Viel} is given by
\begin{equation}
\label{eq:energy1}
	\Omega_Nh^2 \simeq 0.1 \sum\limits_{I=1}^{3}\sum\limits_{\alpha=e,\nu,\tau}\left(\frac{|\Theta_{\alpha I}|^2}{10^{-8}}\right)\left(\frac{M_I}{1 \text{~keV}}\right)^2.
\end{equation}
In particular, the density of $N_1$ particle expressed using the effective mass parameter defined by Eq.\eqref{eq:md:def} is
\begin{equation}
    \label{eq:energy:N1}
    \Omega_{N_1}h^2 \simeq \left( \frac{\sum_{\alpha} (m_D)_{\alpha 1}}{10^{-4}\text{~eV}} \right) \left( \frac{M_1}{1 \text{~keV}} \right).
\end{equation}    
It leads to a restriction from above on $m_D$ summed by flavours
\begin{equation}
    \label{eq:md:dm}
    \overline{(m_D)}_{\text{DM}} = 10^{-5} \left(\frac{M_1}{\text{1~keV}}\right)^{-1} \text{~eV}.
\end{equation}

The excluded area where $\Omega_N > \Omega_{DM} = 0.12$ is shown in Fig.\ref{fig:1} by light-red color. Combining the obtained constraints \eqref{eq:md:xray} and \eqref{eq:md:dm} for the effective mass parameter we get
\begin{equation}
    \label{eq:md:general bound}
    \sum\limits_\alpha (m_D)_{\alpha 1} < \overline{m_D} \equiv \min\left(\overline{(m_D)}_{\rm DM},\overline{(m_D)}_{\rm X-ray}\right)
\end{equation}


\section{On the classification of mixing scenarios}

In the following we consider three possibilities of $\Omega$ matrix parametrization most appropriate to the constraint given by Eq.\eqref{eq:md:general bound},
\begin{itemize}
    \item \textbf{''Fine-tuning'' of mixing} for normal and inverted hierarchies
        \begin{equation}
        \label{eq:omega:ft}
            \Omega^{(FT)}_{\rm NH} =\left(
            \begin{array}{cc}
                1 & 
                \begin{array}{cc}
                0 &  0       
                \end{array}
                \\
                \begin{array}{c}
                0   \\
                0     
                \end{array} & \Omega_{2\times2}
            \end{array}
            \right) \hspace{1cm}
            \Omega^{(FT)}_{\rm IH} =\left(
            \begin{array}{cc}
                \begin{array}{c}
                0   \\
                0     
                \end{array} & \Omega_{2\times2} \\
                1 & 
                \begin{array}{cc}
                0 &  0       
                \end{array}
                \\
            \end{array}
            \right)
    \end{equation}
    where $\Omega_{2\times2}$ is a $2\times 2$ orthogonal matrix. In this form of mixing the constraint is imposed directly on the mass of the lightest active neutrino $m_{\rm lightest}$ ($m_1$ for NH, $m_3$ for IH)
    \begin{equation} \label{finetuning}
        \sum_\alpha (m_D)_{\alpha 1} = \sum_{\alpha,k} |\sqrt{m_k} U_{\alpha1} \delta_{k 1 (k 3)}|^2 = m_{1(3)}.
    \end{equation}
    Here we used the unitarity condition for PMNS matrix assuming that $U \simeq U_{\text{PMNS}}$ up to $\mathcal{O}(\theta^2)$. With such form of mixing matrix there is a "fine-tuning" of mixing that explicitly highlights the non-zero mass of the lightest active neutrino, unlike the scenarios considered in the following, where the small finite numerical value of mass is not so significant. The effective mass parameter $(m_D)_{\alpha 1}$ summed by the flavor index $\alpha$ gives a counterpart of the parameter $U^2_\alpha$ which is used in experimental reconstructions, see Section 6.
\item \textbf{Mixing expressed by the real orthogonal rotation matrix} $\Omega \in SO(3,\mathbb{R})$ with the following parameterization 
    \begin{equation}
    \label{eq:Omega:Euler}
        \Omega =\left(
        \begin{array}{ccc}
            c_2 & -c_3 s_2 & s_2 c_3\\
            c_1 s_2 & c_1 c_2 c_3 - s_1 s_3 & - c_3 s_1 - c_1 c_2 s_3\\
            s_1 s_2 & c_1 s_3 + c_2 c_3 s_1  & c_1 c_3 - c_2 s_1 s_3 
        \end{array}\right)
    \end{equation}
    where $c_j = \cos{\alpha_j}$ and $s_j = \sin{\alpha_j}$, $\alpha_j \in \mathbb{R}$ are Euler angles. Note that the condition \eqref{eq:md:general bound} for matrix \eqref{eq:Omega:Euler} restricts only $\alpha_1$ and $\alpha_2$ angles. Moreover, in this scenario one can take $m_{\rm lightest}=0$. 
\item \textbf{Mixing expressed by the complex special orthogonal matrix $\Omega \in SO(3,\mathbb{C})$} with the same parametrization as given by Eq.\eqref{eq:Omega:Euler} but replacement of $\alpha_j \to \omega_j = \alpha_j + i \beta_j$, $\beta_j \neq 0$. The same as in the previous case, here $m_{\rm lightest}=0$.
\end{itemize}

Possible deviations of $\Omega$ matrix from the form of "fine-tuning" above were analyzed in \cite{jetp}. However, in the following we focus mainly on the form of "fine-tuning" which is consistent with the cosmological constraints in a wide range of HNL dark matter masses, demonstrating also flexibility of the mixing factor in the HNL decays (\ref{eq:width1}) and (\ref{eq:width2}), which can vary due to changes both of $M_1$ and the lightest neutrino mass.

Discussion of ambiguity of the choice of the type of matrix $\Omega$ in the general case can be found in \cite{ibarra2}, the most significant of them are related to the processes of lepton flavor violation \cite{small_lfv} in different sectors of the model. The minimal parametric choice $\Omega=$I corresponds to a special case of "fine tuning" with normal hierarchy, when redundant parameters are not introduced. A similar form for the inverse hierarchy occurs when $\Omega$ is anti-diagonal matrix. In these two cases
 \begin{equation}
 \label{eq:minmix:NH}
    \Theta^{\rm (NH)}_{\min}=\left( \begin{array}{ccc}
    iU_{e1}\sqrt{\frac{m_{1}}{M_{1}}} & iU_{e2}\sqrt{\frac{m_{2}}{M_{2}}} & iU_{e3}\sqrt{\frac{m_{3}}{M_{3}}} \\
    iU_{\mu1}\sqrt{\frac{m_{1}}{M_{1}}} & iU_{\mu2}\sqrt{\frac{m_{2}}{M_{2}}} & iU_{\mu3}\sqrt{\frac{m_{3}}{M_{3}}} \\
    iU_{\tau1}\sqrt{\frac{m_{1}}{M_{1}}} & iU_{\tau2}\sqrt{\frac{m_{2}}{M_{2}}} & iU_{\tau3}\sqrt{\frac{m_{3}}{M_{3}}}
    \end{array} \right),~\text{with}~~~\Omega^{\rm (NH)}_{\min} = \left(\begin{array}{ccc}
        1 & 0 & 0 \\
        0 & 1 & 0 \\
        0 & 0 & 1
    \end{array}\right)
\end{equation}
\begin{equation}
 \label{eq:minmix:IH}
    \Theta^{\rm (IH)}_{\min}=\left( \begin{array}{ccc}
    iU_{e3}\sqrt{\frac{m_{3}}{M_{1}}} & iU_{e2}\sqrt{\frac{m_{2}}{M_{2}}} & iU_{e1}\sqrt{\frac{m_{1}}{M_{3}}} \\
    iU_{\mu3}\sqrt{\frac{m_{3}}{M_{1}}} & iU_{\mu2}\sqrt{\frac{m_{2}}{M_{2}}} & iU_{\mu1}\sqrt{\frac{m_{1}}{M_{3}}} \\
    iU_{\tau3}\sqrt{\frac{m_{3}}{M_{1}}} & iU_{\tau2}\sqrt{\frac{m_{2}}{M_{2}}} & iU_{\tau1}\sqrt{\frac{m_{1}}{M_{3}}}
    \end{array} \right),~\text{with}~~~\Omega^{\rm (IH)}_{\min} = \left(\begin{array}{ccc}
        0 & 0 & 1 \\
        0 & 1 & 0 \\
        1 & 0 & 0
    \end{array}\right)
\end{equation}
where $U_{\alpha I}$ are the elements of $U_{\rm PMNS}$. In further consideration this case of mixing, Eq.\eqref{eq:minmix:NH} or Eq.\eqref{eq:minmix:IH}, is designated as {\it mixing scenario 1}. 

For parametric scenarios in the seesaw type I models which are more interesting for collider phenomenology it is needed to combine very small active neutrino masses of the order of $F^2 v^2/M_{M}$ with moderately heavy HNL, providing observable signals within the LHC and next colliders energy reach, and enhance at the same time small mixing factors of the order of $ \sqrt{{m_\nu}/M_{\rm HNL}}$, providing observable rates at the luminosity frontier. This is achieved either by fine-tuning of the mixing
matrices in a specific scenarios with additional symmetries \cite{smirnov}, or in the framework of Casas-Ibarra diagonalisation with complex-valued parameters. First sort of models gives quasi-Dirac neutrinos processed by the standard calculation technique, which are not fully consistent with the second sort of models beyond the "Dirac limit"{$,$} where evaluations are performed with Majorana fermions. In the latter case $\Omega_{2\times2}$, Eq.\eqref{eq:omega:ft}, is chosen as an element of $SO(2,\mathbb{C})$
\begin{equation} \label{omega22}
    \Omega_{2\times2} (\xi,\omega)= \left(\begin{array}{cc}
        \cos\omega & -\sin\omega \\
        \xi\sin\omega & \xi\cos\omega 
    \end{array}\right)
\end{equation}
In further consideration, this choice is designated as {\it mixing scenario 2}.

 Three new parameters are introduced in (\ref{omega22}), $\xi = \pm1$, ${\rm Re}(\omega) $ and ${\rm Im} (\omega) $. The Dirac limit of scenario 2 has been considered in detail in the literature. Significant enhancements of the collider signals appear with complex-valued $\omega$ parameter which leads to the factors $X_{\omega}=e^{\rm Im}( \omega)$ in the mixing matrix $\Theta$.  Detailed phenomenological analyses of active and sterile neutrino mixing in \cite{asaka-eijima} showed that a phenomenologically consistent hierarchy of mixings $\Theta_{e I}$, $\Theta_{\mu I}$ and $\Theta_{\tau I}$ with suppressed $\Theta_{e I}$ relative to other matrix elements can be achieved in a wide interval of $X_\omega$ independently on the values of HNL masses. Translating the experimental upper bounds on $\Theta_{\alpha I}$ from the shortest possible lifetimes of $N_{2,3}$ from $\pi^\pm$ and $K^\pm$ meson decays into the upper bound on $X_\omega$ one obtains at the HNL mass scale 10$^2$ MeV ${\rm Im}(\omega)=4.5$ for the lifetime of the order of 1 sec and ${\rm Im}(\omega) \sim 7$ for the lifetime of the order of 0.01 sec. Values of ${\rm Im} (\omega)>6-7$ lead to a large mixing parameters of the $W^\pm, Z$-neutrino interactions not consistent with the data. Complex-valued parametrization of $\Omega$ was also used for the study of HNL properties at the TeV scale \cite{smirnov, ibarra3}, see also \cite{alekhin}. 

Extensive literature is devoted to the study of the question of the number of HNL generations. For $\nu$MSM model the case of only two HNL generations in comparison with the case of three generations has been analysed within the cosmological framework in \cite{shaposhnikov1} for arbitrary $\Omega$ and diagonal $M_M$ with the result that the number of HNL generations equal to three is preferred. Neutrino phenomenology for the case of two right-handed neutrinos has been analysed in \cite{ibarraross} where the decoupling limit of the three right-handed neutrino model has been constructed using a specific form form of $\Omega$-matrix in the basis where $M_M$ matrix and the mass matrix of charged standard leptons are diagonal and real with an underlying symmetry for the Yukawa couplings or the elements of $M_M$ (texture zeroes). Constraints on thermal leptogenesis and LFV processes have been found for such a case. In the presence of a sufficiently large number of acceptable cosmological scenarios, we will adhere to the framework of the $\nu$MSM model with dark matter production through active-sterile neutrino mixing, where the mass difference of $N_2$ and $N_3$ is small in comparison with the known mass splittings of the light left-handed neutrino mass states \cite{laine}.

Significant recent reconsideration for the case of two HNL generations in the region of the parameter space corresponding to the mass less than the mass of $K$-meson and performed taking into account the available set of modern data, see \cite{Bondarenko:2021cpc}, is discussed in Section 4 below in connection with the comparison for the case of three HNL generations considered in this paper.
The analysis of lepton universality within scenario 2 in the Dirac limit assuming the $\nu$MSM framework in the approximation of the two-particle decays was performed in \cite{asaka}.

Since in the following the decomposition of the anti-Hermitian matrix $\cal W$, Eq.\eqref{ow}, by powers of $\theta$ to the second order terms is used for the transition to the mass basis of leptons, the question naturally arises about the scope of applicability of such a decomposition and taking into account the ${\cal O}(\theta^3)$ terms of the decomposition (and higher). Within the framework of a non-minimal ${\cal O}(\theta^3)$ decomposition, it is necessary to take into account the terms of the order of ${\cal O}(\theta M_D)$ when \cite{Dubinin:2023yli}
\begin{eqnarray*}
   M_N = (\theta^{-1} - \frac{1}{3} \theta^\dagger) M_D =M_M + \theta^\dagger M_D 
\end{eqnarray*}
whereas, within the framework of the standard mininmal approximation for the see-saw mechanism, it is assumed that $M_N=M_M$. For non-minimal decomposition of the $\cal W$ matrix, the condition must be met
\begin{equation}
    \label{eq:nonmin_omega}
    \Omega^{-1} = \Omega^T + \frac{1}{3}\hat{M}^{-1} (\Omega^{-1})^* \hat{m}, 
\end{equation}
which is a condition for the self-consistency of the diagonalization procedure, taking into account the ${\cal O}(\theta M_D)$ terms. For scenario 2, the mixing matrix, in addition to a small parameter of the order of $\sqrt{m/M}$, contains a potentially large factor of the order of $\Omega^{-1}$, the limited contribution of which must be checked. This issue is discussed in Section 6.

\section{Semileptonic HNL decays}
\begin{figure}[h!]
     \centering
     \begin{subfigure}[t]{0.48\textwidth}
         \centering
                \includegraphics[scale=0.36]{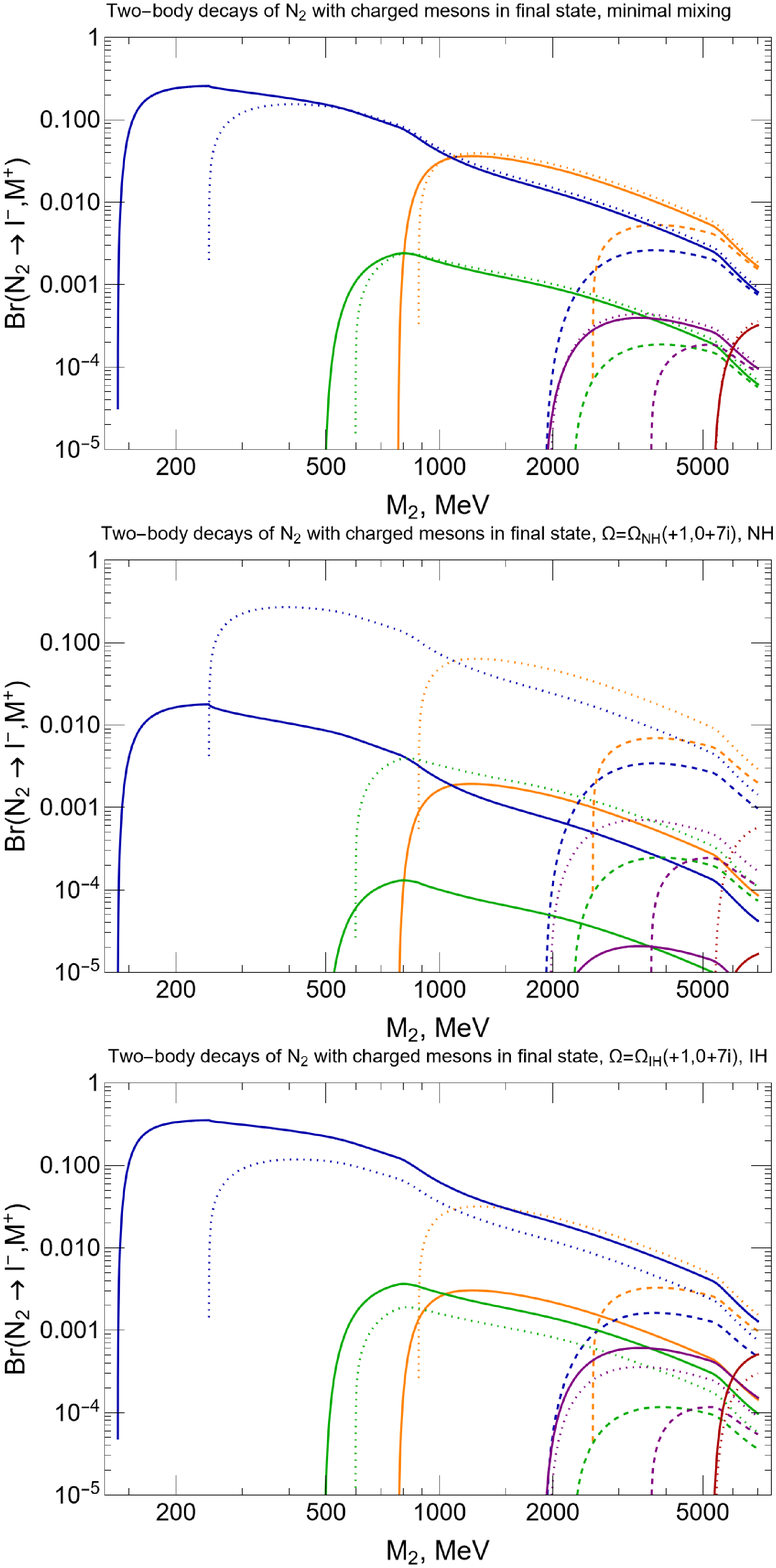}
                \caption{Branching ratios for semileptonic two-body decays of $N_2$ with charged pseudoscalar or vector mesons and charged lepton in the final state.  Solid lines correspond to $e^-$ in final state, long-dashed lines -- to $\mu^-$, dotted lines -- to $\tau^-$. Blue lines -- decay with $\pi^+$, green -- $K^+$, orange -- $\rho^+$, purple -- $D^+$, red -- $B^+$.}
                \label{fig:2:a}
     \end{subfigure}
     \hfill
     \begin{subfigure}[t]{0.48\textwidth}
         \centering
                \includegraphics[scale=0.36]{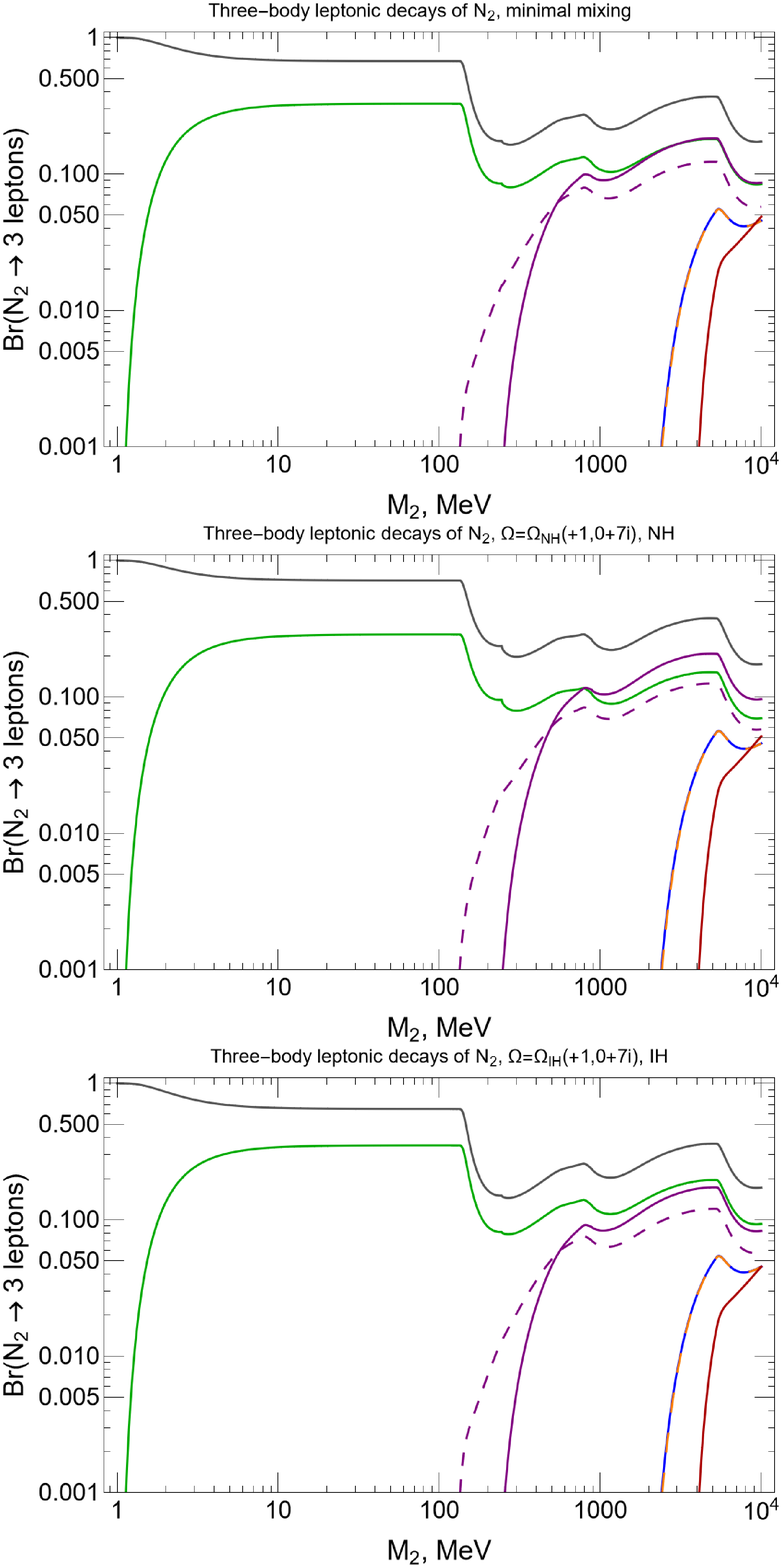}
                \caption{Branching ratios for pure leptonic three-body decays of $N_2$. Gray lines -- $\upnu \upnu \upnu$ decay mode, green lines -- $\upnu e^- e^+$, purple lines -- $\upnu \mu^- \mu^+$, dashed purple lines -- $e^- \mu^+$, dashed blue -- $\upnu e^- \tau^+$, dashed orange -- $\upnu \mu^- \tau^+$, red lines -- $\upnu \tau^-\tau^+$. mode. There is a very weak sensitivity of partial widths in relation to the choice of the mixing scenario.}
                \label{fig:2:b}
     \end{subfigure}
     \caption{Branching ratios for HNL decays. Upper plot --  scenario 1 for both the normal and the inverted hierarchy.  Middle plot -- scenario 2 with $\Omega=\Omega_{NH}(\xi,\omega)$ where $\xi=+1$, ${\rm Re}(\omega)=0$ and ${\rm Im}(\omega)=7$. This parameter set comes from the Big Bang nucleosynthesis (BBN) limits, see Fig.\ref{fig:5}. Bottom plot -- the same as the middle plot but for the case of IH.}
\end{figure}

\begin{figure}[t]
    \centering
    \includegraphics[scale=0.6]{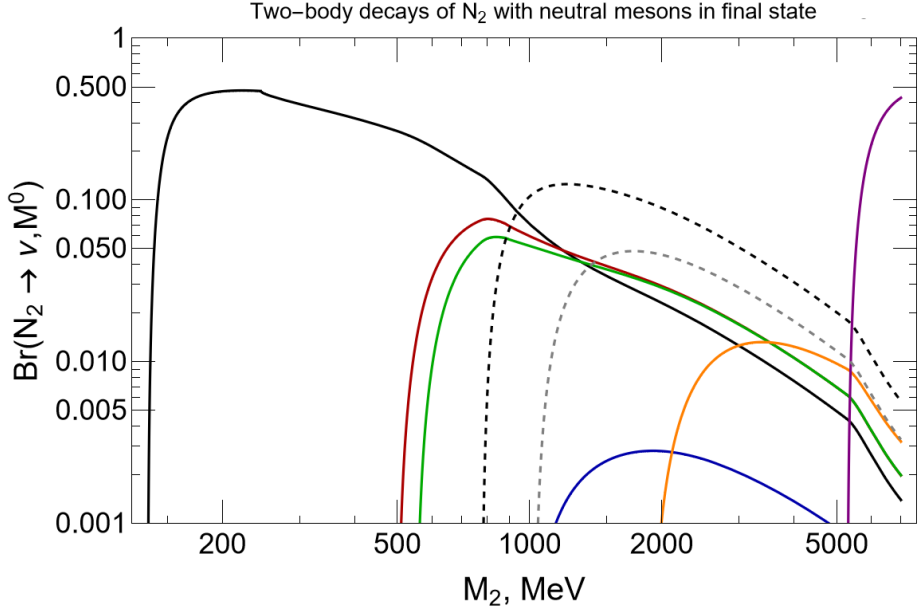}
    \caption{Branching ratios for semileptonic two-body decays of $N_2$ with neutral pseudoscalar or vector mesons and active neutrino in the final state. Curves for NH and IH practically coincide, 
    they are evaluated for the same parametric scenarios as in Fig.\ref{fig:1}. 
    Black lines -- decay with $\pi^0$, red -- $K^0$, green -- $\eta$, blue -- $\eta^\prime$, orange -- $D^0$, purple -- $B^0$, dashed black -- $\rho^0$, dashed gray -- $\varphi$. This type of decay is not sensitive to flavor and mixing scenarios 1 and 2 for both hierarchies do not have significant differences.}
            \label{fig:3}        
\end{figure}

HNL decays with meson in the final state can be divided in four groups \cite{Gorbunov-Shaposh}:
\begin{itemize}
    \item pseudoscalar neutral meson $M^0_{ps} = \pi^0,\eta, \eta^\prime, K^0, D^0, B^0$ in the final state with the decay width
        \begin{eqnarray} 
        \label{decay:ps0}
            \Gamma (N_I \to \nu_\alpha h^0_{ps}) = \frac{G_F^2 M_I^3}{32\pi} f_{M^0_{ps}} |\Theta_{\alpha I}|^2  \left(1 - x_{M}^2\right)^2,
    \end{eqnarray}
    \item pseudoscalar charged meson $M^\pm_{ps} = \pi^\pm,K^\pm,D^\pm,D_s^\pm,B^\pm,B_s^\pm,B_c^\pm$            
    \begin{equation}
        \label{decay:ps+}
            \Gamma (N_I \to l^-_\alpha M^+_{ps}) = \frac{G_F^2 M_I^3}{16\pi} f_{M^+_{ps}} |V_{q q^\prime}|^2 |\Theta_{\alpha I}|^2 \left(\left(1-x_\alpha^2\right)^2 - x_{M}^2 \left(1+x_\alpha^2\right)\right) \lambda^{\frac{1}{2}} (1,x_{M}^2,x_\alpha^2),
    \end{equation}
    \item vector neutral meson $h^0_v=\rho^0,\varphi,\omega$
    \begin{equation}
         \label{decay:vec0}
            \Gamma (N_I \to \nu_\alpha h^0_v) = \frac{G_F^2 M_I^3}{16\pi} \frac{g^2_{M^0_v}}{m^2_{M^0_v}}|V_{qq^\prime}|^2|\Theta_{\alpha I}|^2 (1 + 2x^2_{M}) \left(1 - x_{M}^2\right)^2,
    \end{equation}
    \item vector charged meson $M^\pm_v=\rho^\pm$ in the final state with the decay width
        \begin{multline}
        \label{decay:v+}
            \Gamma (N_I \to l^{-}_\alpha M^+_v) = \frac{G_F^2 M_I^3}{32\pi} \kappa_M \frac{g^2_{M^0_v}} {m^2_{M^0_v}}|\Theta_{\alpha I}|^2 \times \\
            \times \left(\left(1-x_\alpha^2\right)^2 + x_{M}^2 \left(1+x_\alpha^2\right) - 2x_{M}^4\right)\lambda^{\frac{1}{2}} (1,x_{M}^2,x_\alpha^2).
        \end{multline}
\end{itemize}
where $G_F$ is the Fermi constant, $V_{qq^\prime}$ is the Cabibbo-Kobayashi-Maskawa (CKM) matrix element, $f_M$ and $g_M$ are the corresponding meson decay constants \cite{Gorbunov-Shaposh}, $x_M=m_M/M_{\rm HNL}$, $x_\alpha=m_{\alpha}/M_{\rm HNL}$, $m_\alpha$ is the mass of charged lepton $l_\alpha$, $\lambda$ is two-particle kinematic function,
$\kappa_M$ is an additional dimensionless correction factor for hadronic matrix element $\langle 0 | J_\mu^Z | M_v^0\rangle$ (see \cite{Gorbunov-Shaposh} for details). Values of $\kappa_M$ are given in Table \ref{tab:kappa}.
\begin{table}[h!]
   \centering
    \begin{tabular}{| m{1cm} |m{2cm}|m{2cm}|m{2cm}|m{2cm}|}
        \hline
        $M_v^0$ & $\rho^0$ & $\omega$ & $\phi$ & $J/\psi$ \\
        \hline
        $\kappa_M$ & $1-2s_w^2$ & $\frac{4}{3} s_W^2$ & $\frac{4}{3} s_W^2 - 1$ & $1 - \frac{8}{3}s_W^2$ \\
        \hline
    \end{tabular}
    \caption{Values for correction factor $\kappa_M$ in Eq.\eqref{decay:v+} from \cite{Gorbunov-Shaposh}, $s_W=\sin \theta_W$ sine of the Weinberg angle}
    \label{tab:kappa}
\end{table}
These two-particle widths, see Fig.\ref{fig:2:a} and Fig.\ref{fig:3}, introduce essential contributions to the total $N_2$ width starting from the $\pi^0$ threshold.

\section{Leptonic HNL decays}
In this section we calculate squared amplitudes and HNL decay widths for the four-fermion effective interaction terms when 
$M_{1,2,3} \ll M_W,M_Z$. Three different decay amplitudes with respect to the mixing factors can be distinguished, $N_I \to \upnu_i,\upnu_j, \upnu_j;$  $N_I \to \upnu_i,l^+_\beta,l^-_\beta$ and $N_I \to \upnu_i,l^+_\alpha, l^-_\beta$. Three-particle decay widths  are calculated symbolically ($p = k_1+k_2+k_3$, $p^2=M_I^2$, $k_i^2=m_i^2$, i=1,2,3) keeping all masses of leptons nonzero by integrating in invariant variables over the Dalitz plot. The details of these calculations are given in Appendix B.
In the model under consideration, all six neutrinos are Majorana fermions, which requires careful determination of the signs of the interference terms. See in this connection \cite{denner} for the {\it fermion flow} technique for diagrams with Majorana fermions, or \cite{haber} for the case of generic basis of $\gamma$-matrices. Three cases for amplitudes take place:\\ 

{\bf Case 1}. Three active neutrino mass states appear in the final state: $N_I(p) \to \upnu_{i}(k_1), \upnu_j(k_2), \upnu_j(k_3)$, $i,j=1,2,3$.  Insofar as the approximation $\mathcal{O}(\theta^2)$ is used, the PMNS matrix can be operated as a unitary one $(U^\dagger U)=\delta_{ij}$ and there are no $Z\upnu_j \upnu_{k\neq j}$ vertices. Neglecting the active neutrino masses for the sake of clarity we get the squared amplitude of the following structure
    \begin{equation*}
    \label{decay:MM:3nu:1}
        |M_{3\upnu}|^2 = 64 G_F^2 |(U^\dagger\Theta)_{iI}|^2~\left[ (p k_3) (k_1 k_2) + (p k_2) (k_1 k_3)\right]
    \end{equation*}
Details of the calculations can be found in Appendix~A.
    \begin{figure}[h!]
        \begin{center}
            \begin{fmfgraph*}(120,120)
                \fmfstraight
                \fmfleft{i1,i2,id1,id2,i3,i4,i5}
                \fmfright{o1,o2,od1,od2,o3,o4,o5}
                \fmf{plain,tension=1.5,label=$N_I(p)$, label.side=right, label.dist=0.1cm}{v2,i4}
                \fmf{plain,label=$\upnu_i(k_1)$, label.side=right, label.dist=0.1cm}{o4,v2}
                \fmffreeze
                \fmf{plain}{o1,v1,o3}
                \fmf{plain,tension=3,label=$\upnu_j(k_3)$, label.side=right, label.dist=0.1cm}{o1,v1}
                \fmf{plain,label=$\upnu_j(k_2)$, label.side=right, label.dist=0.1cm}{v1,o3}
                \fmf{photon, tension=2,label=$Z$, label.side=right, label.dist=0.2cm}{v2,v1}
            \end{fmfgraph*}
        \end{center}
        \vspace{1em}
        \begin{center}
            Diagram for the case 1. HNL decay into three active Majorana neutrinos.
        \end{center}
    \end{figure}

The mixing factor $|(U^\dagger \Theta)_{iI}|^2$ is reduced taking the sum over the active neutrino mass states
    \begin{multline*}
      \sum \limits_{i}|(U^\dagger \Theta)_{iI}|^2 = \sum \limits_{i} \sum \limits_{\alpha, \beta} U^*_{\alpha i} \Theta_{\alpha I} U_{\beta i} \Theta^*_{\beta I} =  \sum \limits_{\alpha, \beta} (UU^\dagger)_{\alpha \beta} \Theta_{\alpha I} \Theta^*_{\beta I} = \sum \limits_{\alpha, \beta} \delta_{\alpha \beta} \Theta_{\alpha I} \Theta^*_{\beta I} = \sum \limits_{\alpha}|\Theta_{\alpha I}|^2
    \end{multline*}
and the decay width  
    \begin{equation}
    \label{decay:3nu}
    \Gamma(N_I \to \sum \limits_{i,~j} \upnu_i,\upnu_j, \upnu_j) = \frac{G_F^2 M_I^5}{96 \pi^3} \sum \limits_{\alpha=e,\mu,\tau} |\Theta_{\alpha I}|^2.
    \end{equation}
    
{\bf Case 2}. Two leptons of different flavors and a neutrino mass state appear in the final state: $N_I(p) \to \upnu_j(k), l_{\alpha \neq \beta}^+(p_1), l_\beta^-(p_2)$. Neglecting the interference term between diagrams with intermediate $W^+$ and  $W^-$, 
\begin{figure}[h!]
    \begin{center}
        \begin{fmfgraph*}(100,100)
            \fmfstraight
            \fmfleft{i1,i2,id1,id2,i3,i4,i5}
            \fmfright{o1,o2,od1,od2,o3,o4,o5}
            \fmf{plain,tension=1.5,label=$N_I(p)$, label.side=right, label.dist=0.1cm}{v2,i4}
            \fmf{fermion,label=$l_\alpha^+(p_1)$, label.side=right, label.dist=0.1cm}{o4,v2}
            \fmffreeze
            \fmf{plain}{o1,v1,o3}
            \fmf{plain,tension=3,label=$\upnu_k(k)$, label.side=right, label.dist=0.1cm}{o1,v1}
            \fmf{fermion,label=$l_\beta^-(p_2)$, label.side=right, label.dist=0.1cm}{v1,o3}
            \fmf{photon, tension=2,label=$W^+$, label.side=right, label.dist=0.2cm}{v2,v1}
        \end{fmfgraph*}
            \hspace{3cm}
        \begin{fmfgraph*}(100,100)
            \fmfstraight
            \fmfleft{i1,i2,id1,id2,i3,i4,i5}
            \fmfright{o1,o2,od1,od2,o3,o4,o5}
            \fmf{plain,tension=1.5,label=$N_I(p)$, label.side=right, label.dist=0.1cm}{v2,i4}
            \fmf{fermion,label=$l_\beta^-(p_2)$, label.side=left, label.dist=0.1cm}{v2,o4}
            \fmffreeze
            \fmf{plain}{o1,v1,o3}
            \fmf{plain,tension=3,label=$\upnu_k(k)$, label.side=right, label.dist=0.1cm}{o1,v1}
            \fmf{fermion,label=$l_\alpha^+(p_1)$, label.side=left, label.dist=0.1cm}{o3,v1}
            \fmf{photon, tension=2,label=$W^-$, label.side=right, label.dist=0.2cm}{v2,v1}
        \end{fmfgraph*}
    \end{center}
    \vspace{1em}
    \begin{center}
        Diagrams for the case 2. HNL decay into Dirac charged lepton and antilepton of different flavors associated with one active Majorana neutrino
    \end{center}
\end{figure}
\begin{equation*}
    |M_W|^2 = 128 G_F^2 \left[|\Theta_{\alpha I}|^2 |U_{\beta i}|^2 (p p_2)(p_1 k) + |\Theta_{\beta I}|^2 |U_{\alpha i}|^2 (p p_1)(p_2 k)\right].
\end{equation*}
After summing by the active neutrino mass states and using $UU^\dagger=$I due to approximate unitarity of PMNS matrix, the decay width takes the form
\begin{equation} 
    \label{eq:width:case3}
    \Gamma(N_I \to \sum \limits_{i=1,2,3}\upnu_i~l^{+}_{\alpha \neq \beta} l^{-}_\beta) = \frac{G_F^2 M_I^5}{192\pi^3} \left(|\Theta_{\alpha I}|^2  + |\Theta_{\beta I}|^2 \right) \mathcal{G}(r_\alpha,r_\beta),
\end{equation}
where 
    \begin{eqnarray*}
        \mathcal{G}(x,y) &=& (1 - 7 x - 7 x^2 + x^3  + 12 x y - 7 y - 7 y^2  + y^3 - 7 x^2 y - 7 x y^2)R + \\
        && + 12(y^2 + x^2y^2 - 2x^2) \ln\left(\frac{1+x-y+R}{2}\right) + 12x^2(1-y^2) \ln(\frac{1}{x}) + \\
        && + 12y^2(1-x^2) \ln\left(\frac{1-x-y+R}{1-x+y-R}\right),~~~~ R = \sqrt{1 -2x + x^2 -2y + y^2 - 2xy} 
    \end{eqnarray*}
In the limiting case $m_{e} \ll m_{\mu}$, $m_{\mu} \ll m_{\tau}$ Eq.\eqref{eq:width:case3} is reduced to
\begin{multline}
     \Gamma(N_I \to \sum \limits_{i=1,2,3}\upnu_i~l^{+}_{\alpha \neq \beta} l^{-}_\beta)\Big|_{m_{\beta} \to 0} = \Gamma(N_I \to \sum \limits_{i=1,2,3}\upnu_i~l^{+}_{\alpha \neq \beta} l^{-}_\beta)\Big|_{m_{\alpha} \to 0} = \\ =\frac{G_F^2 M_I^5}{192\pi^3}  \left(|\Theta_{\alpha I}|^2 + |\Theta_{\beta I}|^2\right) \left(1 - 8 r  + 8r^3 - r^4 - 12r^2\ln(r)\right)
\end{multline}
where $r=\frac{m^2}{M_I^2}$, $m=max\{m_\alpha, m_\beta\}$. This result is the same as the one obtained in \cite{Gorbunov-Shaposh}.

Note that if one assumes that the active neutrinos are Dirac fermions, then there is no interference between diagrams with $W^+$ and $W^-$ because they correspond to different final states $\overline{\upnu} l^+ l^-$ and $\upnu l^{+} l^{-}$. Discussion of the approaches to evaluations for Dirac and Majorana fermions can be found in \cite{dune},  
where the Dirac limit is used for the neutrinos. Beyond the Dirac limit, the interference term between diagrams with Majorana neutrinos vanishes if the active neutrino mass is taken to be zero. 
To illustrate the suitability of approximations of this kind it is useful to calculate the decay width with interference term using a simplified amplitude neglecting the $\frac{m}{M_I}$ power terms, and integrating over the triangle Dalitz plot
\begin{equation}
    \Gamma(N_J \to \sum \limits_{k=1}^3\upnu_k l_\alpha^+ l_\beta^-) = \frac{G_F^2 M_J^5}{192 \pi^3} \left(|\Theta_{\alpha J}|^2 + |\Theta_{\beta J}|^2 - \frac{4}{M_J}\sum\limits_{k=1}^3 m_k Re\{\Theta_{\alpha J} \Theta_{\beta J}^* U_{\beta k}^* U_{\alpha k}\}\right)
\end{equation}
One can observe that the interference term, suppressed by the mass ratio, may be not small in the case of scenario 2 with huge mixing factors of the order of $e^{\rm Im}(\omega)$ and for HNL masses at the eV scale (see the discussion of such scales in \cite{alekhin}). The sign of interference term depends on the sign of ${\rm Re}(\omega)$.

\begin{figure}
    \centering
    \begin{subfigure}[t]{0.48\textwidth}
        \includegraphics[scale=0.6]{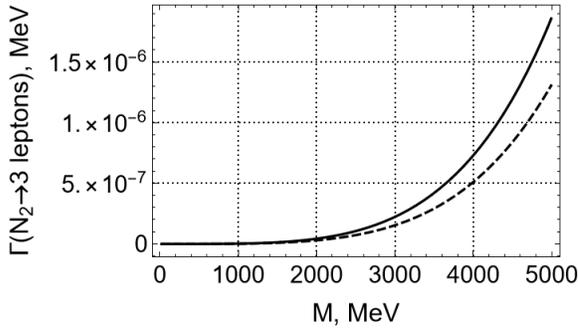}
    \caption{Partial widths of 3-body leptonic HNL decays $N\to l_\alpha,l_\alpha,\upnu$ evaluated in the Dirac limit (dashed line) and using Eq.(\ref{gamma_case3}) (solid line).}
    \label{fig:4:a}
    \end{subfigure}
\hfill
    \begin{subfigure}[t]{0.48\textwidth}
    \includegraphics[scale=0.6]{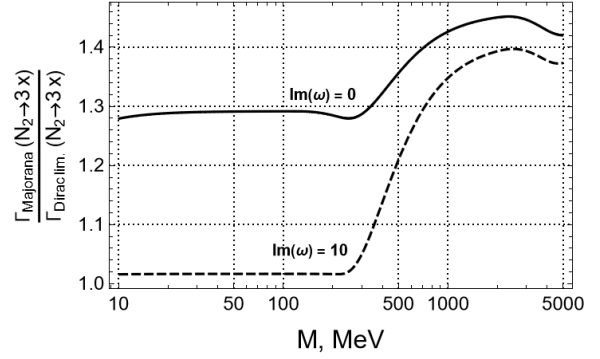}
    \caption{Ratios of partial widths for 3-body leptonic HNL decays $N\to l_\alpha,l_\alpha,\upnu$ evaluated in the Dirac limit and using Eq.(\ref{gamma_case3}) for ${\rm Im}(\omega)=0$ (solid line) and ${\rm Im}(\omega)=10$ (dased line).}
    \label{fig:4:b}
    \end{subfigure}
    \caption{Comparison of the HNL leptonic decay widths calculating in the Dirac limit and in model with 3 Majorana active neutrinos}
    \label{fig:4}
\end{figure}

{\bf Case 3}. Two leptons of the same flavor and a neutrino mass state appear in the final state: $N_I(p) \to \upnu_j(k), l_{\alpha}^+(p_1), l_\alpha^-(p_2)$. The amplitude contains both charged and neutral currents and includes in this case three interfering diagrams
    \begin{center}
        \begin{fmfgraph*}(100,100)
            \fmfstraight
            \fmfleft{i1,i2,id1,id2,i3,i4,i5}
            \fmfright{o1,o2,od1,od2,o3,o4,o5}
            \fmf{plain,tension=1.5,label=$N_I(p)$, label.side=right, label.dist=0.1cm}{v2,i4}
            \fmf{fermion,label=$l_\alpha^+(p_1)$, label.side=right, label.dist=0.1cm}{o4,v2}
            \fmffreeze
            \fmf{plain}{o1,v1,o3}
            \fmf{plain,tension=3,label=$\upnu_k(k)$, label.side=right, label.dist=0.1cm}{o1,v1}
            \fmf{fermion,label=$l_\beta^-(p_2)$, label.side=right, label.dist=0.1cm}{v1,o3}
            \fmf{photon, tension=2,label=$W^+$, label.side=right, label.dist=0.2cm}{v2,v1}
        \end{fmfgraph*}
            \hspace{1.5cm}
        \begin{fmfgraph*}(100,100)
            \fmfstraight
            \fmfleft{i1,i2,id1,id2,i3,i4,i5}
            \fmfright{o1,o2,od1,od2,o3,o4,o5}
            \fmf{plain,tension=1.5,label=$N_I(p)$, label.side=right, label.dist=0.1cm}{v2,i4}
            \fmf{fermion,label=$l_\beta^-(p_2)$, label.side=left, label.dist=0.1cm}{v2,o4}
            \fmffreeze
            \fmf{plain}{o1,v1,o3}
            \fmf{plain,tension=3,label=$\upnu_k(k)$, label.side=right, label.dist=0.1cm}{o1,v1}
            \fmf{fermion,label=$l_\alpha^+(p_1)$, label.side=left, label.dist=0.1cm}{o3,v1}
            \fmf{photon, tension=2,label=$W^-$, label.side=right, label.dist=0.2cm}{v2,v1}
        \end{fmfgraph*}
            \hspace{1.5cm}
        \begin{fmfgraph*}(100,100)
            \fmfstraight
            \fmfleft{i1,i2,id1,id2,i3,i4,i5}
            \fmfright{o1,o2,od1,od2,o3,o4,o5}
            \fmf{plain,tension=1.5,label=$N_I(p)$, label.side=right, label.dist=0.1cm}{v2,i4}
            \fmf{plain,label=$\upnu_k(k)$, label.side=right, label.dist=0.1cm}{o4,v2}
            \fmffreeze
            \fmf{fermion}{o1,v1,o3}
            \fmf{fermion,tension=3,label=$l_\alpha^+(p_1)$, label.side=right, label.dist=0.2cm}{o1,v1}
            \fmf{fermion,label=$l_\alpha^-(p_2)$, label.side=right, label.dist=0.2cm}{v1,o3}
            \fmf{photon, tension=2,label=$Z$, label.side=right, label.dist=0.2cm}{v2,v1}
        \end{fmfgraph*}
    \end{center}
    \begin{center}
        Diagrams for the case 3. HNL decay into Dirac charged lepton and antilepton with the same flavor and one active Majorana neutrino.
    \end{center}
The mixing factor appearing in the squared amplitude is reduced by summing over the states of active neutrinos\footnote{The squared amplitude is not summed up by neutrino mass states $\upnu_{1,2,3}$, see details of calculation in Appendix A.}
    \begin{multline}
       \sum \limits_{i}\Theta_{\alpha I} U^*_{\alpha i} (U^\dagger\Theta)_{iI}^* = \sum \limits_{i,\beta} \Theta_{\alpha I} U^*_{\alpha i} U_{\beta i} \Theta_{\beta I} = \sum \limits_{\beta} (UU^\dagger)_{\alpha\beta} \Theta_{\alpha I}\Theta^*_{\beta I} = \\
       =\sum \limits_{\beta} \delta_{\alpha\beta} \Theta_{\alpha I}\Theta^*_{\beta I} =  |\Theta_{\alpha I}|^2.
    \end{multline}
Integration of the amplitude 
    \begin{eqnarray*}
        \sum \limits_i |M_{\rm WZ}|^2 &=& 128 G_F^2 \left[\left(\mathcal{C}_1^2 + \mathcal{C}_2^2\right) \sum \limits_\beta |\Theta_{\beta I}|^2 + \left(1 - 2\mathcal{C}_1\right) |\Theta_{\alpha I}|^2 \right] ~\left[(p p_2)(p_1 k)+(p p_1)(p_2 k)\right] + \\
        && + 128 G_F^2 \left[\left(4\mathcal{C}_1\mathcal{C}_2\right)\sum \limits_\beta |\Theta_{\beta I}|^2 - 4\mathcal{C}_2 |\Theta_{\alpha I}|^2\right] ~ \frac{m_\alpha^2}{2} (p k)
    \end{eqnarray*}
gives the decay width 
\begin{eqnarray} \label{m2wz}
    \nonumber \label{gamma_case3}
    \Gamma(N_I \to \sum \limits_{i=1,2,3}\upnu_i~l^{+}_\alpha l^{-}_\alpha) = \frac{G_F^2 M_I^5}{96\pi^3} &\Bigg(& \left[\left(\mathcal{C}_1^2 + \mathcal{C}_2^2\right) \sum \limits_\beta |\Theta_{\beta I}|^2 + \left(1 - 2\mathcal{C}_1\right) |\Theta_{\alpha I}|^2 \right] \mathcal{F}_1(r) + \\ 
    && + 4\left[\mathcal{C}_1\mathcal{C}_2 \sum \limits_\beta |\Theta_{\beta I}|^2 - \mathcal{C}_2 |\Theta_{\alpha I}|^2\right] ~\mathcal{F}_2(r) \Bigg)
    \end{eqnarray}
where $\mathcal{C}_1 = s_W^2-\frac{1}{2}$, $\mathcal{C}_2 = s_W^2$, $r=\frac{m_{\alpha}^2}{M_I^2}$,
\begin{eqnarray*} 
    \mathcal{F}_1(r) &=&  \left(1 - 14 r - 2 r^2 - 12 r^3\right) \sqrt{1-4 r}~+~ 12r^2\left(1-r^2\right)\ln \left(\frac{1-3r+(1-r)\sqrt{1-4r}}{r(1-\sqrt{1-4r})}\right), \\
    \mathcal{F}_2(r) &=& (2r+10r^2-12r^3)\sqrt{1-4r}~-~ (6r^2 - 12r^3 + 12r^4) \ln\left(\frac{1-3r+(1-r)\sqrt{1-4r}}{r(1-\sqrt{1-4r})}\right).
\end{eqnarray*}
In the charged lepton massless limit $r=0$, $\mathcal{F}_1(r)=$1, $\mathcal{F}_2(r)=$0 and the factor $-2\mathcal{C}_1$ in front of the interference term is positive. Comparison of the decay width calculated using Eq.(\ref{gamma_case3}) and the decay width in the Dirac limit is shown in Fig.\ref{fig:4:a},b. The difference in three-particle widths, caused by the opposite signs of the interference terms in Eq.(\ref{gamma_case3}) compared with the Dirac limit, can be several times, however, the main contribution to the total HNL width made by two-particle modes is significantly greater than the three-particle modes.

{\bf Case of HNL in final state}. Decay channels with HNL in the final state $N_{2,3} \to N_1 + ...$ are suppressed by the factor of $\sim \mathcal{O}(\Theta^2)$ in comparison with the channels described above. They are insignificant for the following analysis.
              
        \begin{figure}
                \centering
            \begin{subfigure}[t]{0.48\textwidth}
                \includegraphics[scale=0.4]{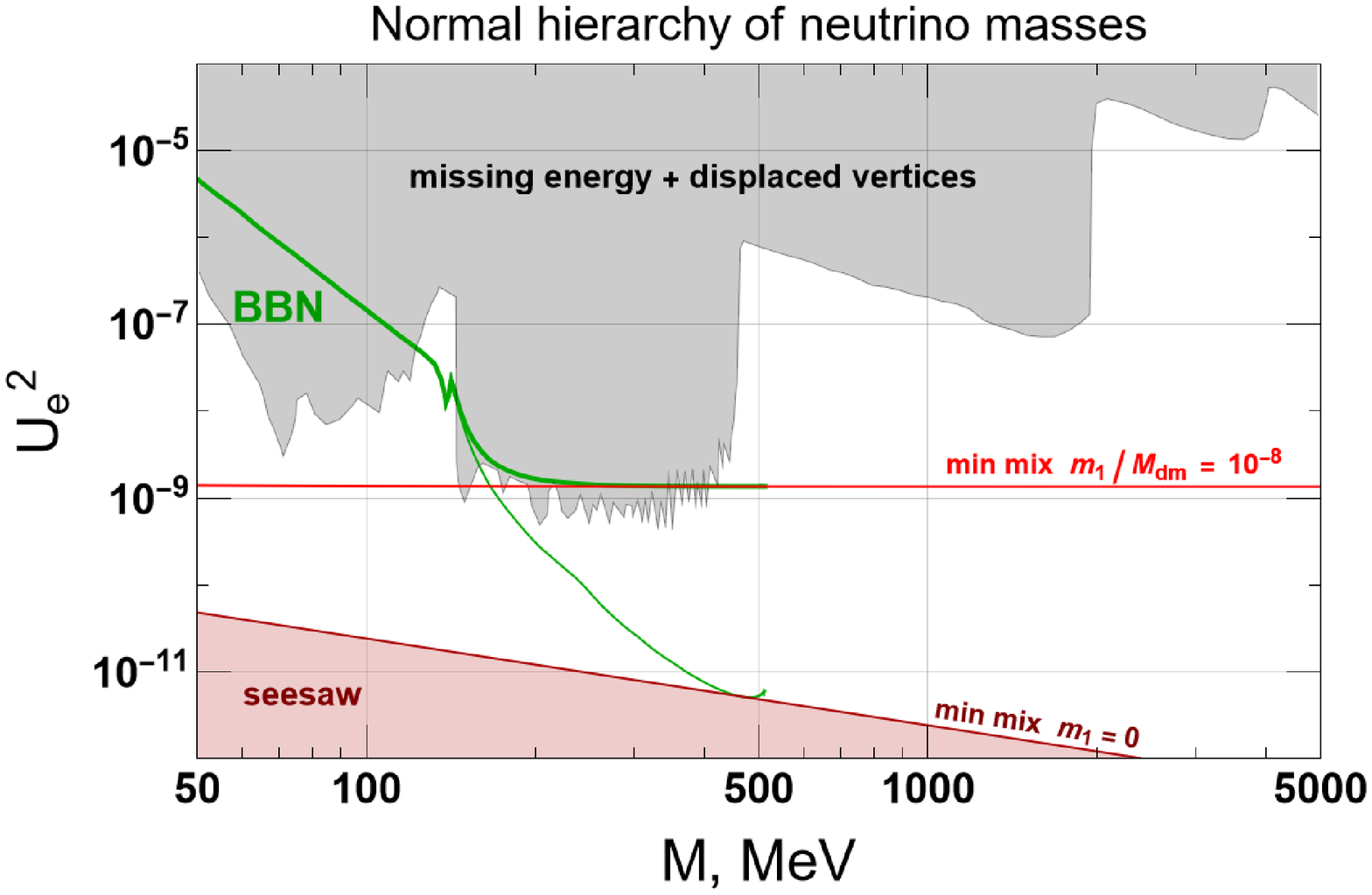}
                \includegraphics[scale=0.4]{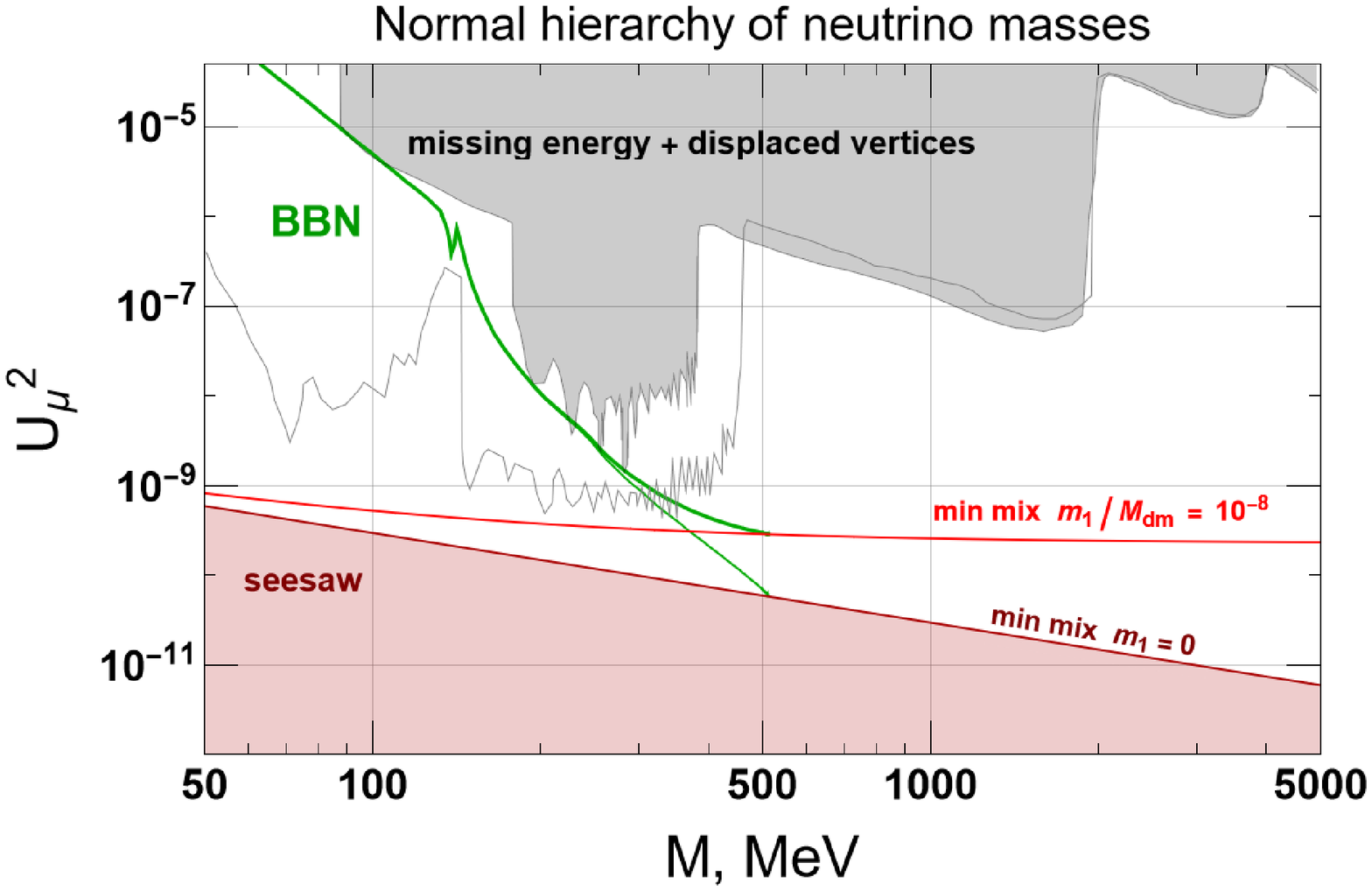}
            \caption{NH}
            \label{fig5:nh}
            \end{subfigure}
        \hfill
            \begin{subfigure}[t]{0.48\textwidth}
                \includegraphics[scale=0.4]{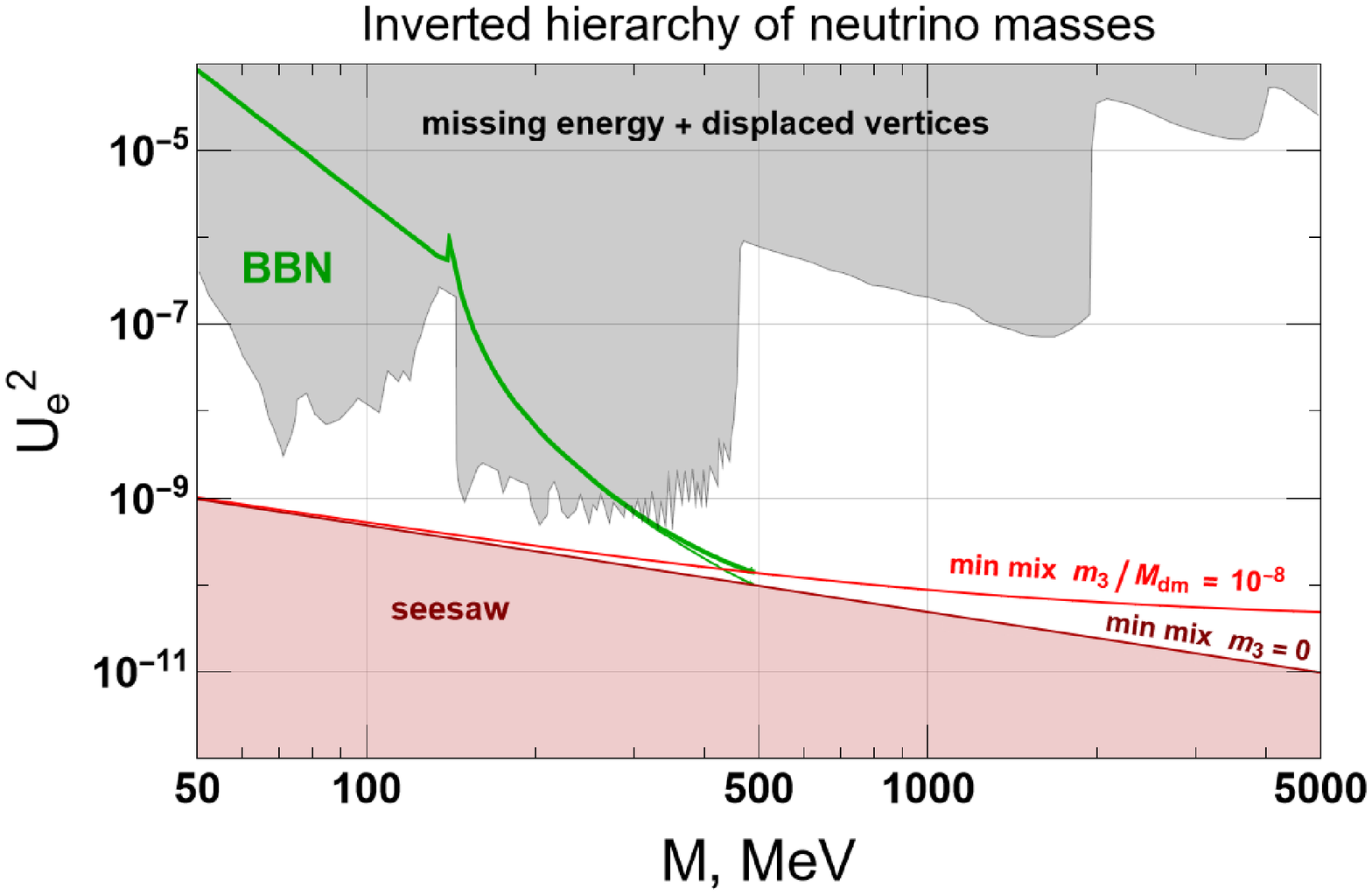}
                \includegraphics[scale=0.4]{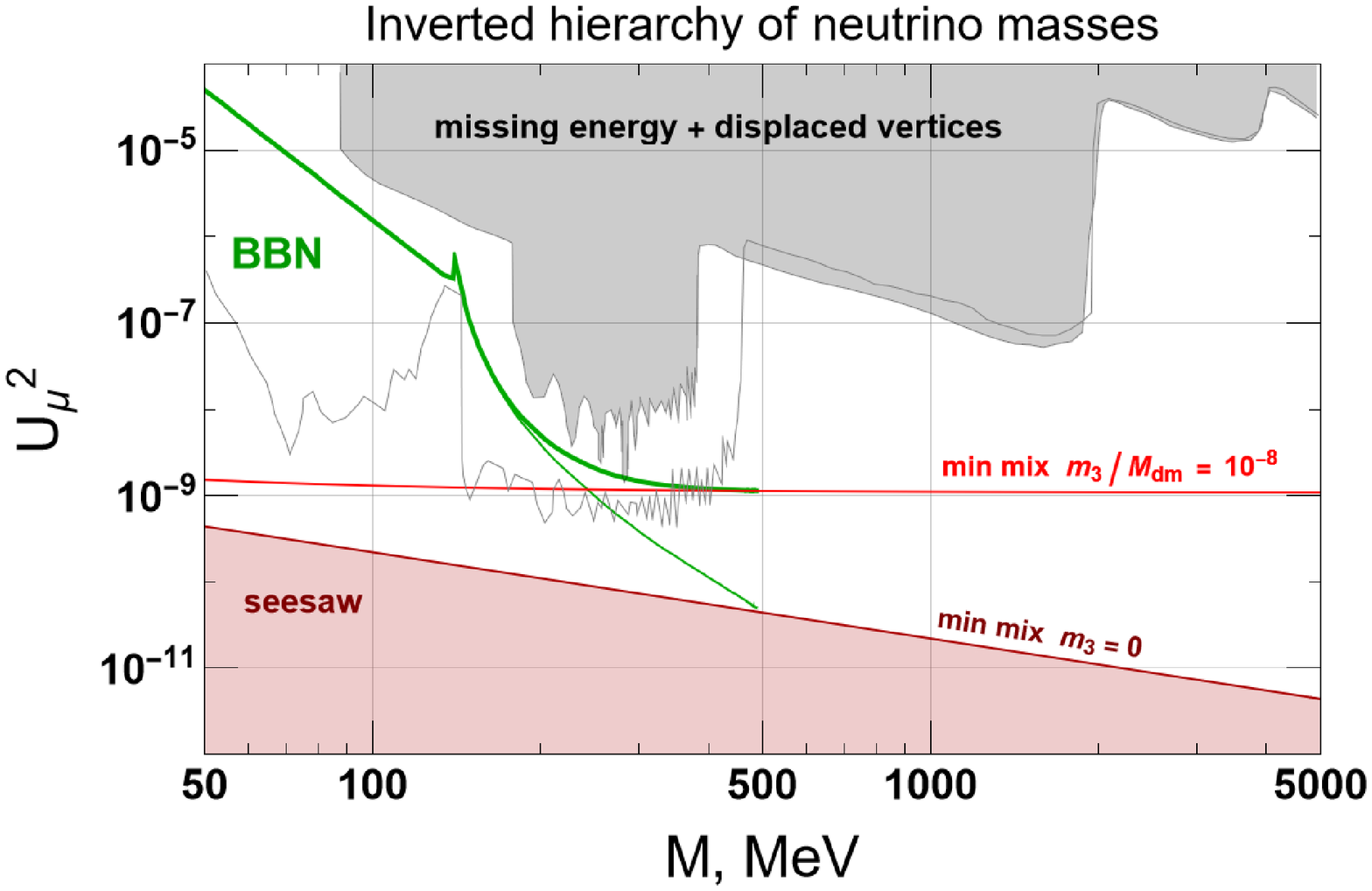}
            \caption{IH}
            \label{fig5:ih}
            \end{subfigure}
            \centering
           
            \caption{
            Constraints on $U_e^2$ and $U_\mu^2$ for various HNL masses. Green lines are the lower bounds derived from the BBN lifetime limit, left panels -- NH, right panels -- IH; thin green lines are for the simplified model with two HNL's, thick green lines -- for the full model with three HNLs and the lightest active neutrino mass $m_1/M_{DM}$ or $m_3/M_{DM}$ of the order of $10^{-8}$. Excluded domains in gray color correspond to the upper limits from two types of accelerator experiments: experiments with the missing energy reconstruction (TRIUMPH \cite{triumph}, PIENU \cite{pienu} ($\pi$ decay), NA62 \cite{na62}, E949 \cite{e949}, KEK \cite{kek} ($K$ decay)) and experiments with the displaced vertices (DELPHI \cite{delphi}, PS-191 \cite{ps191}, CHARM \cite{charm}, NuTeV \cite{nutev} ), the division of the gray region into subdomains corresponding to different accelerator experiments can be found in \cite{Bondarenko:2021cpc}.
            Seesaw bound contours of $U_e^2$ and $U_\mu^2$ in the simplified case of two HNLs and/or the case of three HNLs with $m_{1(3)}=0$ are shown by red and dark red lines. 
            }
            \label{fig:5}
        \end{figure}
        

\section{Constraints on the mixing parameters for $N_2$ and $N_3$ }

In both scenarios defined in Section 3 the contribution of $N_1$, the DM particle, to the lepton universality parameter is small and the degree of lepton universality violation (LUV) depends on $N_2$ and $N_3$.

\subsection{Upper bounds from accelerator experiments}
There are experimental restrictions for phenomenological parameters defined as
\begin{eqnarray}
    U^2_\alpha &=& \sum\limits_{I=1}^{3} |\Theta_{\alpha I}|^2\\
    U^2_I &=& \sum\limits_{\alpha=e,\mu,\tau} |\Theta_{\alpha I}|^2,\\
    U^2 &=& \sum\limits_\alpha \sum\limits_I |\Theta_{\alpha I}|^2 = Tr(\Theta^\dagger \Theta).
\end{eqnarray}
For the decay channels $\pi, K \to e, \mu + E_{miss}$ the missing energy is reconstructed in the experiments TRIUMPH \cite{triumph}, PIENU \cite{pienu}, NA62 \cite{na62}, E949 \cite{e949} and KEK \cite{kek}. In the experiments DELPHI \cite{delphi}, PS191 \cite{ps191}, CHARM \cite{charm}, NuTeV \cite{nutev} an identification of HNL decay displaced vertices is carried out. A combination of bounds from these two types of experiments taken from \cite{Bondarenko:2021cpc} is shown in Fig.\ref{fig:5}.

Taking into account the valuable remark in review \cite{alekhin} regarding the use of the so-called "model-independent approach" in the analysis of data from various experiments\footnote{assumption of the independence of the mixing parameter from the HNL mass seems to be quite strong even within the framework of consideration with one generation of leptons, see for example \cite{gronau}. Neutrino oscillations can be described with at least two generations of HNL's, two generations are involved in the low-scale leptogenesis.}, we note the need for careful translation when bringing the results to a common denominator. In the general case when the mass and the mixing parameter are not independent variables, the exclusion contours are subject to dependence on the field-theoretic model of the expansion of the lepton sector. The partial probabilities of HNL decays in the model under consideration with six Majorana neutrinos differ from the corresponding probabilities in the model with active Dirac neutrinos and Majorana sterile neutrinos. These deviations can be significant and may lead to some quantitative displacement of the exclusion contours, although a qualitative correspondence will take place.

Significant development beyond the model-independent approach has been performed in \cite{Bondarenko:2021cpc} where two HNL generations with degenerate masses have been introduced with the 3$\times$2 $\Omega$ matrix analogous to Eq.\eqref{omega22}. In this case $m_1=$0 in the active neutrino mass matrix for NH and $m_3=$0 for IH.

\subsection{Lower bound for BBN}
Cosmological considerations imposing restrictions on the lifetime of $N_2$ and $N_3$ on the level of $\tau_{N_{2,3}}<$~0.1 -- 1 sec \cite{dolgov} were recently improved in \cite{improved-bbn} giving the minimal level of 0.02 sec. These bounds are model dependent and obtained in the framework of a rather specific scenarios of the Big Bang nucleosynthesis. A simplified estimate for BBN lifetime limit is used in the following evaluations 
\begin{equation}
    \label{eq:bbn_lifetime}
    \tau_{\rm HNL} < \tau_{\rm BBN} = \Big\{\begin{array}{c}
    0.02~\text{sec},~ M_{\rm HNL} > 140~\text{MeV}\\
    0.1~\text{sec},~ M_{\rm HNL} < 140~\text{MeV}
    \end{array}
\end{equation}

\begin{figure}
    \centering
        \includegraphics[scale=0.5]{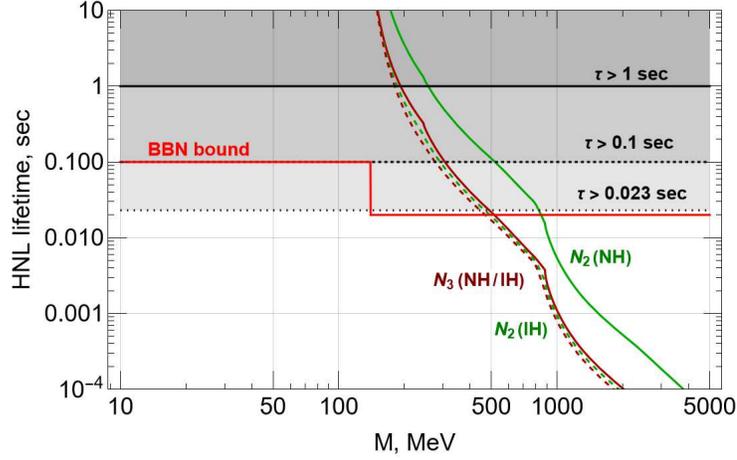}
        \caption{The lifetimes of $N_2$ (green lines) and $N_3$ (dark red lines) as a function of their masses $M_2 = M_3$ in the case of normal (solid lines) and inverted (dashed lines) hierarchies, scenario 1. Red line correspond to BBN constraints \eqref{eq:bbn_lifetime}, which is taken from \cite{improved-bbn}. $N_2$ contour for the case of normal active neutrino hierarchy practically coincides with $N_3$ contours for both hierarchies.}
        \label{fig:6}
\end{figure}

In the scenario 1 framework the minimal mixing matrix \eqref{eq:minmix:NH} does not contain redundant parameters, so the lifetime dependence on the HNL mass is unambiguous, see Fig.\ref{fig:6}. 

For scenario 2, it is necessary to take into account the constraint for $\Omega$ matrix elements following from the self-consistency condition of the model with Casas-Ibarra diagonalization extended to the cubic terms in the decomposition of the $\cal W$ matrix, see Eq.\eqref{omega}. The exponential factor $e^{2{\rm Im}(\omega)}$ may give a huge increase of the mixing parameters $|\Theta_{\alpha2}|^2$ and $|\Theta_{\alpha3}|^2$
\begin{eqnarray} \label{asymptotic}
    \left|\Theta^{\rm (NH)}_{\alpha2(3)}\right|^2\bigg|_{{\rm Im}(\omega)>1}  &\simeq&  \frac{e^{2 {\rm Im}(\omega)}}{4 M_{2(3)}} \left|\sqrt{m_2} U_{\alpha2} + i \xi \sqrt{m_3} U_{\alpha3}\right|^2 + \mathcal{O}\left(\dfrac{\sqrt{\Delta m_{atm}^2}}{M_{2(3)}}\right),
    \\
    \left|\Theta^{\rm (IH)}_{\alpha2(3)}\right|^2\bigg|_{{\rm Im}(\omega)>1}  &\simeq&  \frac{e^{2 {\rm Im}(\omega)}}{4 M_{2(3)}} \left|\xi\sqrt{m_2} U_{\alpha2} - i \sqrt{m_1} U_{\alpha1}\right|^2 + \mathcal{O}\left(\dfrac{\sqrt{\Delta m_{atm}^2}}{M_{2(3)}}\right)
\end{eqnarray}

\begin{figure}
    \centering
        \begin{subfigure}[t]{0.48\textwidth}
            \includegraphics[scale=0.5]{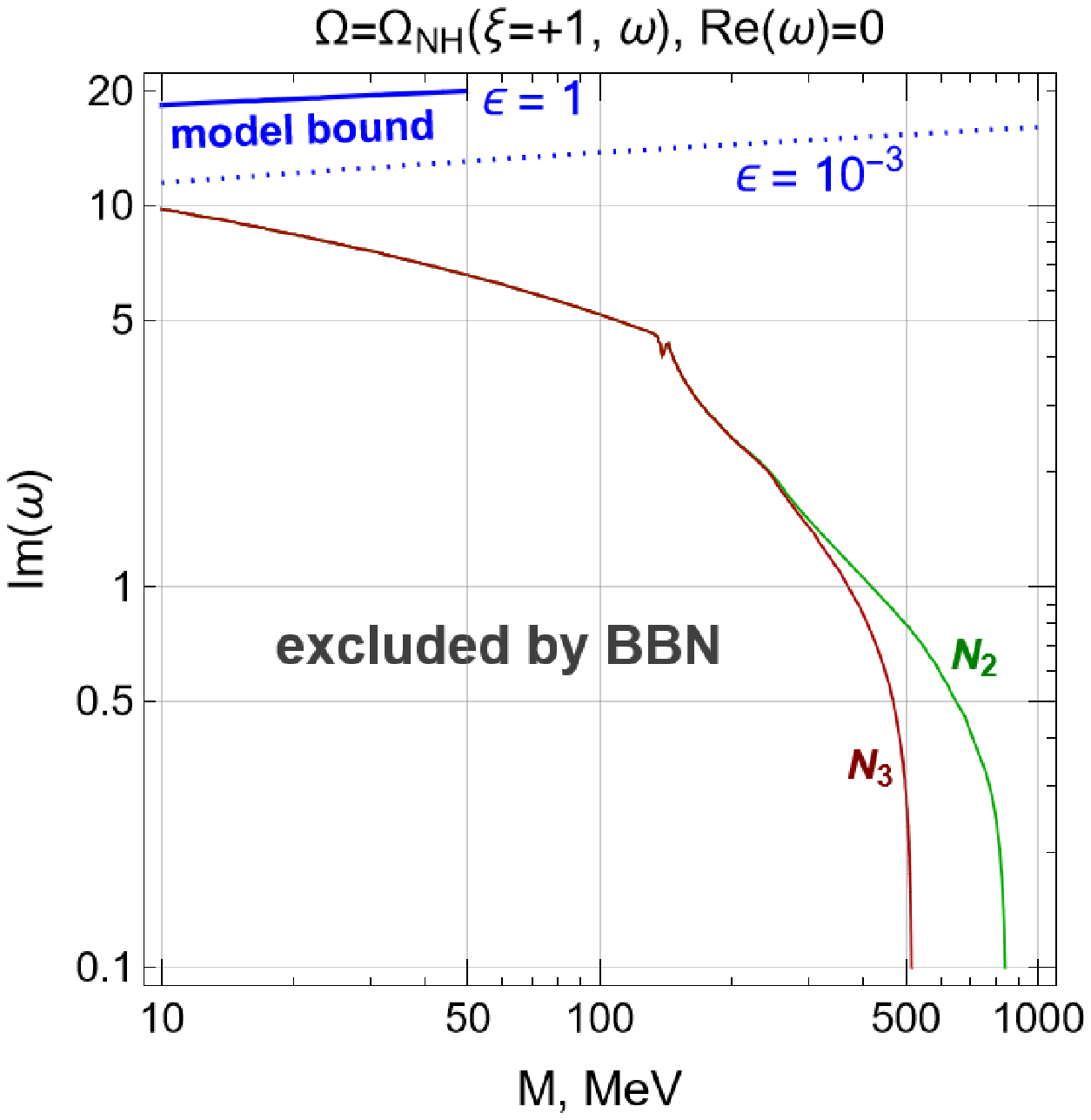}
            \caption{NH}
            \label{fig6:nh}
        \end{subfigure}
        \hfill
        \begin{subfigure}[t]{0.48\textwidth}
            \includegraphics[scale=0.5]{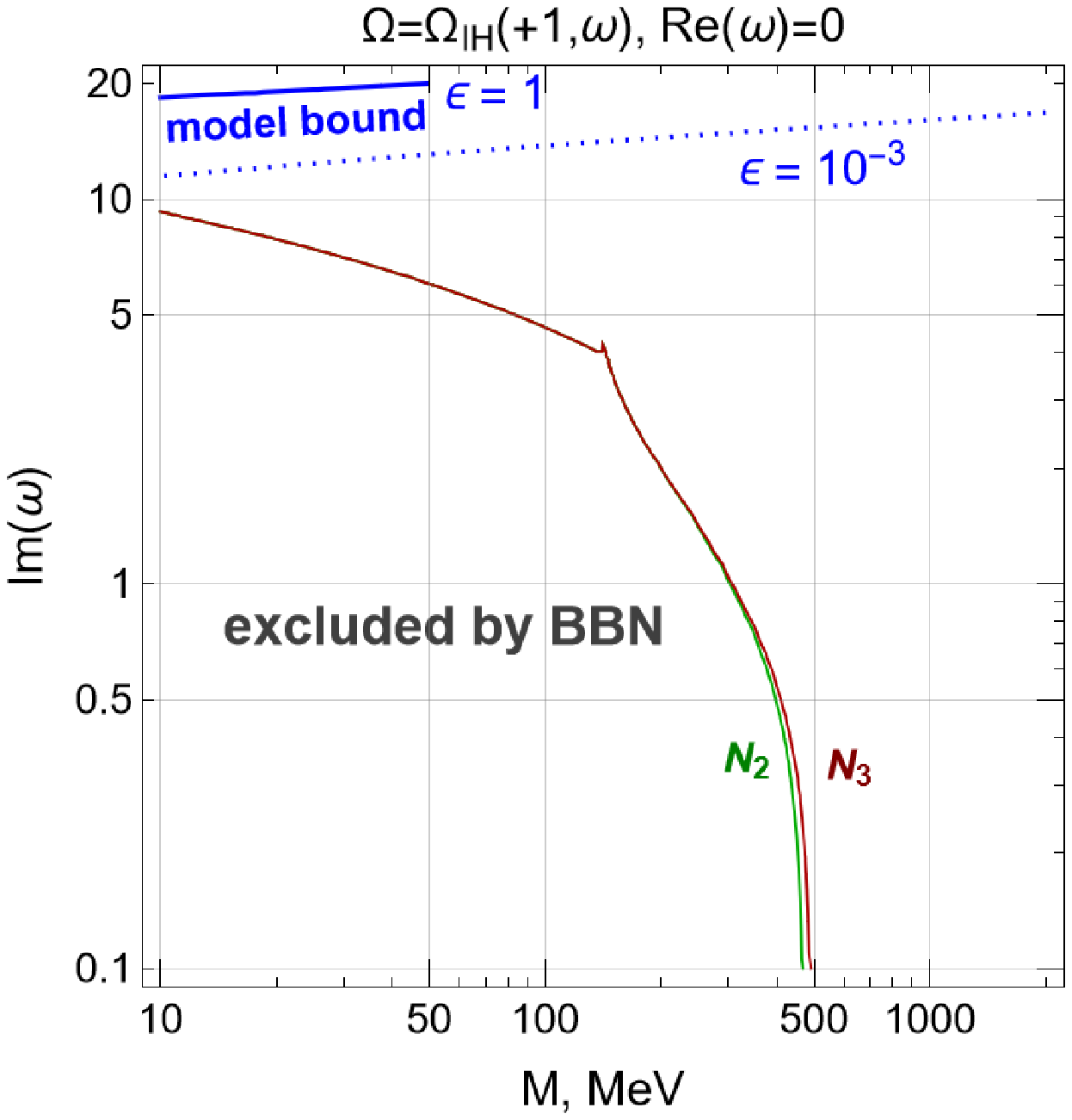}
            \caption{IH}
            \label{fig6:ih}
        \end{subfigure}
    \centering
    \caption{The BBN constraints for HNL lifetime (green line for $N_2$ and dark red line for $N_3$) as a function of its mass in scenario 2 with following parameterization of mixing $\Omega=\Omega_{\rm NH}(\xi=+1,\omega)$ for NH (a) and $\Omega=\Omega_{\rm IH}(\xi=+1,\omega)$ for IH (b) with ${\rm Re}(\omega)=0$. Blue lines correspond to the restriction from above on the ${\rm Im}(\omega)$ parameter \eqref{eq:nonmin_omega:bound} with $\epsilon=1$ (solid line) and $\epsilon=10^{-3}$ (dotted line). The allowed parameter domains for corresponding HNL are above the lines marked by $N_2$ (green line) and $N_3$ (dark red line) and below the blue solid line.}
    \label{fig:7}
\end{figure}

\begin{figure}[t]
            \centering
            \includegraphics[scale=0.7]{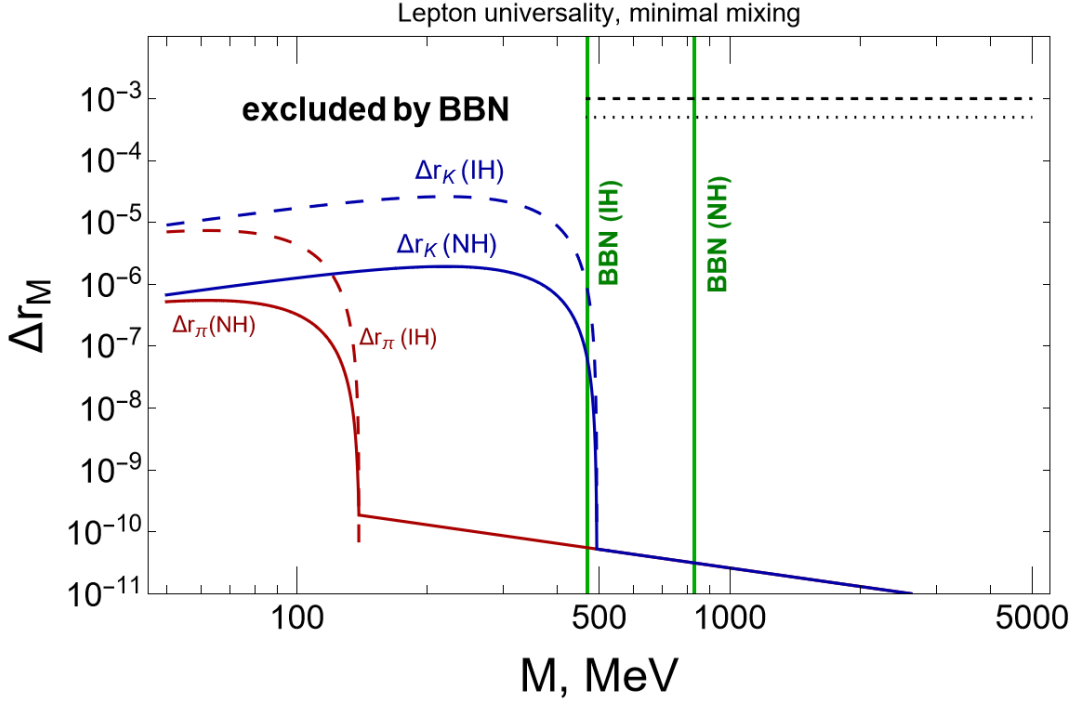}
            \caption{Lepton universality parameters $\Delta r_{\pi}$, red lines, and $\Delta r_K$, blue lines, as a function of the HNL mass in the case of quasidegenerate $M_2\simeq M_3 \equiv M$ for scenario 1. First generation HNL $N_1$ with the mass $M_1\simeq 5$ keV is the dark matter particle. Vertical green lines are the lower bounds from BBN lifetime constraints for IH (left line) and NH (right line). Horizontal dashed lines corrspond to experimental values of LUV parameters: $\Delta r_\pi = 5\cdot10^{-4}$ and $\Delta r_K = 10^{-3}$, see \cite{pdg}.}
            \label{fig:8}
        \end{figure}

Using the constraint of Eq.\eqref{eq:nonmin_omega} for the orthogonal matrix $\Omega$, one arrives to
\begin{equation}
\label{eq:nonmin_omega:epsilon}
    \frac{1}{3}\hat{M}^{-1} (\Omega^{-1})^* \hat{m} \Big|_{{\rm Im}(\omega) > 1} \simeq 0, ~~\text{so}~~ \frac{m_{2,3}}{M_{2,3}}e^{{\rm Im}(\omega)} \equiv \epsilon \ll 1, 
\end{equation}
and for $m_{2,3} \simeq O(0.1~\text{eV})$ the inequality must hold
\begin{equation}
\label{eq:nonmin_omega:bound}
    {\rm Im}(\omega) \lesssim 16.1 + \ln{\left[\epsilon \left(\frac{M_{2,3}}{1~\text{MeV}}\right)\right]}.
\end{equation}
In this scenario first the allowed lifetime domain on the ${\rm Im}(\omega)-M$ plane is found, see Fig.\ref{fig:7}, which is then translated to the allowed domain for the lepton universality parameter $\Delta r_M$, see Fig.\ref{fig:8} and  Fig.\ref{fig:luv_kaon}. In the phenomenological analysis, in the following we use the commonly accepted denomination for the model parameter space
\begin{equation}
    U_\alpha^2 = \sum \limits_{I=1}^{3} |\Theta_{\alpha I}|^2 = \Bigg\{\begin{array}{c}
         \frac{m_{1}}{M_1}|U_{\alpha1}|^2+|\Theta^{(NH)}_{\alpha2}|^2+|\Theta^{(NH)}_{\alpha3}|^2,~~~\text{NH}\\
         \frac{m_3}{M_1}|U_{\alpha3}|^2 + |\Theta^{(IH)}_{\alpha2}|^2+|\Theta^{(IH)}_{\alpha3}|^2,~~~~\text{IH}
    \end{array} 
\end{equation}
where $m_{1(3)}$ is the mass of lightest active neutrino for normal (inverted) hierarchy. The case of simplified model with two HNL generations, see \cite{Bondarenko:2021cpc}, can be reproduced in the limiting case of the model under consideration when $m_{1(3)} \to 0$ is taken, which is valid for the range of HNL masses where the lower bound of $U_\alpha^2 \gg |\Theta_{\alpha1}|^2$ at fixed non-zero $m_1$ or $m_3$. 
        
\subsection{Lower bound for seesaw}
In the limiting case of ${\rm Im}(\omega)\to 0$ which is the transition to the real-valued $\Omega$ matrix, it is necessary to take into account all terms, since they become comparable in the order of magnitude 
\begin{equation}
    \label{eq:theta:real:NH}
    \lim_{Im(\omega)\to0}\Theta^{\rm (NH)} = \left(
    \begin{array}{ccc}
     \sqrt{\frac{m_1}{M_1}} U_{e1} & \frac{\sqrt{m_2} U_{e2} \cos \phi+ \xi\sqrt{m_3}  U_{e3} \sin \phi}{\sqrt{M_2}} & \frac{ \xi\sqrt{m_3}  U_{e3} \cos \phi-\sqrt{m_2} U_{e2} \sin \phi}{\sqrt{M_3}} \\
    \sqrt{\frac{m_1}{M_1}} U_{\mu1} & \frac{\sqrt{m_2} U_{\mu2} \cos \phi+ \xi\sqrt{m_3}  U_{\mu3} \sin \phi}{\sqrt{M_2}} & \frac{ \xi\sqrt{m_3}  U_{\mu3} \cos \phi-\sqrt{m_2} U_{\mu2} \sin \phi}{\sqrt{M_3}} \\
     \sqrt{\frac{m_1}{M_1}} U_{\tau1} & \frac{\sqrt{m_2} U_{\tau2} \cos \phi+ \xi\sqrt{m_3}  U_{\tau3} \sin \phi}{\sqrt{M_2}} & \frac{ \xi\sqrt{m_3}  U_{\tau3} \cos \phi-\sqrt{m_2} U_{\tau2} \sin \phi}{\sqrt{M_3}} \\
    \end{array}
    \right),
\end{equation}
\begin{equation}
    \label{eq:theta:real:IH}
    \lim_{Im(\omega)\to0}\Theta^{\rm (IH)} = \left(
    \begin{array}{ccc}
     \sqrt{\frac{m_3}{M_1}} U_{e3} & \frac{\sqrt{m_1} U_{e1} \cos \phi+ \xi\sqrt{m_2}  U_{e2} \sin \phi}{\sqrt{M_2}} & \frac{ \xi\sqrt{m_2}  U_{e2} \cos \phi-\sqrt{m_1} U_{e1} \sin \phi}{\sqrt{M_3}} \\
     \sqrt{\frac{m_3}{M_1}} U_{\mu3} & \frac{\sqrt{m_1} U_{\mu1} \cos \phi+ \xi\sqrt{m_2}  U_{\mu2} \sin \phi}{\sqrt{M_2}} & \frac{ \xi\sqrt{m_2}  U_{\mu2} \cos \phi-\sqrt{m_1} U_{\mu1} \sin \phi}{\sqrt{M_3}} \\
     \sqrt{\frac{m_3}{M_1}} U_{\tau3} & \frac{\sqrt{m_1} U_{\tau1} \cos \phi+ \xi\sqrt{m_2}  U_{\tau2} \sin \phi}{\sqrt{M_2}} & \frac{ \xi\sqrt{m_2}  U_{\tau2} \cos \phi-\sqrt{m_1} U_{\tau1} \sin \phi}{\sqrt{M_3}} \\
    \end{array}
    \right),
\end{equation}
where $\phi = {\rm Re}(\omega)$. If zero mass of the lightest active neutrino is taken, $m_{1}=0$ (NH) or $m_3 = 0$ (IH), then the values of mixing matrix elements form the "seesaw bound" as it is called in the existing literature, see for example \cite{Bondarenko:2021cpc}. Note that in the case of nearly degenerate $M_2$ and $M_3$ inherent to $\nu$MSM-like model which we adhere, there is no $\phi$-dependence of the mixing parameter $U_\alpha^2$ in the case of ${\rm Im}(\omega) = 0$ when $\Theta$-matrix has the form \eqref{eq:theta:real:NH} or \eqref{eq:theta:real:IH} (due to $\sin^2 \phi + \cos^2 \phi =$1). \textit{Seesaw bound} is inroduced to mark a minimal possible value of phenomenological parameters \footnote{The more accurate calculation of $\min (U_\alpha^2)$ shows that the real minimal value occurs when ${\rm Im}(\omega) \neq 0$ and differs for different $\alpha$. However, this boundary shift is insignificant for our consideration. For example, the seesaw bound decreases by less than one half for $U_e^2$ when ${\rm Im}(\omega) \simeq \pm 0.5$ for $\xi=\pm1$.}, in particular for $U_\alpha^2$ we can write
\begin{equation}
    \label{eq:uu:min}
    U^2_{\alpha~\min} = \sum\limits_{\alpha=e,\mu,\tau} |(\Theta_{\min})_{\alpha 1}|^2.
\end{equation}
It is appropriate to call this parameter at $m_{1}=0$ or $m_3=0$ as the "absolute seesaw bound" for NH or IH, and if we choose a non-zero mass of the lightest active neutrino, then to call \eqref{eq:uu:min} as just a "seesaw bound". These two bounds coincide when $\frac{m_1}{M_1} \ll \frac{m_2}{M_2}, \frac{m_3}{M_3}$. For $\nu$MSM-like model with $m_{\rm light} \sim 10^{-5}$ eV, $M_1 \sim 1-10$ keV and $M_2 \simeq M_3 \sim 1- 10^4$ MeV such inequatity is not respected, since
\begin{equation}
    \frac{m_{\rm light}}{M_1} \sim 10^{-8} - 10^{-9} ~~~\text{and}~~~ \frac{m_{2,3}}{M} \sim 10^{-8} - 10^{-13}
\end{equation}
so the mixing component of Dark Matter HNL $N_1$ becomes the dominant term of $U^2_\alpha$, or at least the same order of magnitude term as the other terms for $M_1\sim~1$ keV. The difference in bounds is illustrated in Fig.\ref{fig:5}.

\section{Restrictions on the lepton universality parameter in the decays of $\pi^\pm$ and $K^\pm$ mesons}
Limitations on the lepton universality parameter are imposed by the restrictions on the $N_2$ and $N_3$ lifetime from the Big Bang scenario and experimental restrictions from meson decays.

Although the calculations were performed taking into account all the abovementioned decay modes, it is instructive to write out simple formulas for the case of leptonic two-particle decays. In the effective four-fermion approximation the width of the scalar meson $M=\pi^\pm,K^\pm$
\begin{equation}
\label{eq:Mdecay:HNL}
    \Gamma(M^+ \to l_\alpha^+, N_I) = \frac{G_F^2 f_M^2}{4 \pi} |\Theta_{\alpha I}|^2 m_M^4 \lambda^{1/2}(1, r_I, r_\alpha) \left[r_I + r_\alpha - (r_I-r_\alpha)^2\right]
\end{equation}
where $G_F$ is Fermi constant and $f_M$ is meson formfactor, $\alpha = e,\mu,\tau$ (only channels allowed by energy-momentum conservation are admitted), $I=1,2,3$ number of HNL generation, $j=1,2,3$ active neutrino mass states; 
$r_{\alpha} = m^2_{\alpha}/m^2_M$, $r_I= M_I^2/m_M^2$.\\
In the case of only one active neutrino in the final state the mass corrections are neglected and
\begin{equation}
\label{eq:Mdecay:Nu}
    \Gamma(M^+ \to l_\alpha^+, \upnu_j) = \frac{G_F^2 f_M^2}{4 \pi} |U_{\alpha i}|^2 m_M^4 r_\alpha(1-r_\alpha)^2. 
\end{equation}
Convenient variable to analyse BSM deviations from lepton universality is the ratio \cite{shrock1}
\begin{equation}
\label{eq:rm:1}
    R_M = \frac{\sum_i \Gamma(M^+ \to e^+ \upnu_i) + \sum_j \Gamma(M^+ \to e^+ N_j)}{\sum_i \Gamma(M^+ \to \mu^+ \upnu_i) + \sum_j \Gamma(M^+ \to \mu^+ N_j)}
\end{equation}
or its derivative demonstrating the deviation of the ratio from zero
\begin{equation}
\label{eq:rm:2}
    \Delta r_M \equiv \frac{R_M}{R^{SM}_M} \, -1 = \frac{\sum_i |U_{ei}|^2 + \sum_I|\Theta_{eI}|^2 G^{M}_{eI}}{\sum_i |U_{\mu i}|^2 + \sum_I|\Theta_{\mu I}|^2 G^{M}_{\mu I}}\, -1.
\end{equation}
where masses of active neutrinos are neglected and only decays which are allowed kinematically $M_I<m_M-m_\alpha$ are accounted for. If  $M_I>m_M-m_\alpha$ then $G_{\alpha I}^M = 0$. The SM parameter and the BSM parameter are
\begin{eqnarray}
    R_M^{SM} &=& \frac{r_\mu(1-r_\mu)^2}{r_e(1-r_e)^2}, \nonumber \\
    G^{M}_{\alpha I} &=& \frac{\lambda^{1/2}(1,r_I,r_\alpha)\left[r_I + r_\alpha - (r_\alpha - r_I)^2 \right]}{r_\alpha(1-r_\alpha)^2}. \nonumber 
\end{eqnarray}

Using the unitarity condition for the full $6 \times 6$ mixing matrix
\begin{equation}\label{eq:full_unit_cond}
    \sum \limits_{i=1}^3 |U_{\alpha i}|^2 + \sum \limits_{I=1}^3|\Theta_{\alpha I}|^2 = 1
\end{equation}
one can rewrite \eqref{eq:rm:2} in the form 
\begin{equation}
\label{eq:rm:3}
    \Delta r_M =\frac{1 + \sum\limits_{I=1}^3|\Theta_{eI}|^2 (G^M_{eI} - 1)}{1 + \sum\limits_{I=1}^3|\Theta_{\mu I}|^2(G^M_{\mu I} - 1)} - 1
\end{equation}
\subsection{Numerical analysis}
Taking into account the combined restrictions (see Fig.\ref{fig:5}), the allowed values of the mass are $M>430$ MeV for $U_e^2$-bounds and $M>300$ MeV for $U_\mu^2$-bounds. Consequently, in the case of $\pi$-decay, the lepton universality is violated due to the non-unitarity of the PMNS matrix and
\begin{equation}
    \label{eq:Delta_r:pi}
    \Delta r_\pi = \frac{1+(G^\pi_{e1}-1)|\Theta_{e1}|^2 - \left(U_e^2 - |\Theta_{e1}|^2\right)}{1+(G^\pi_{\mu1}-1)|\Theta_{\mu1}|^2 - \left(U_\mu^2 - |\Theta_{\mu1}|^2\right)}-1.
\end{equation}
For $K$-meson decay the function $G^K_{eJ}$ is nonzero in an allowed range $430 \text{~MeV} < M < 497$~MeV. Moreover, if $M_2$ and $M_3$ are quasidegenerate, then $G^K_{\alpha 2}\equiv G^K_{\alpha 3}$ and 
\begin{equation}
    \label{eq:Delta_r:K}
    \Delta r_K = \frac{1+(G^K_{e1}-1)|\Theta_{e1}|^2 + (G^K_{e2}-1)\left(U_e^2 - |\Theta_{e1}|^2\right)}{1+(G^K_{\mu1}-1)|\Theta_{\mu 1}|^2 + (G^K_{\mu2}-1)\left(U_\mu^2 - |\Theta_{\mu1}|^2\right)}-1.
\end{equation}
For the dark matter fermion $N_1$ we take $M_1=5$ keV. The values for the contribution of terms with $m_1=10^{-5}$ eV are given in Table \ref{table:3}. Components of the mixing matrix are 
\begin{equation*}
    |\Theta_{e1}|^2 = \Big\{ \begin{array}{c}
         1.35\cdot 10^{-9}  ~~\text{for NH} \\
        4\cdot 10^{-11}  ~~\text{for IH}
    \end{array} \hspace{1cm} |\Theta_{\mu1}|^2 = \Big\{ 
    \begin{array}{c}
         2.2\cdot 10^{-10} ~~\text{for NH}\\
         1.08 \cdot 10^{-9} ~~\text{for IH}
    \end{array}
\end{equation*} 
so terms $(G^M_{\alpha1}-1)|\Theta_{\alpha 1}|^2 \sim 10^{-14}-10^{-21}$ can be neglected in comparison with other terms. 

\begin{figure}[!h]
    \centering
    \includegraphics{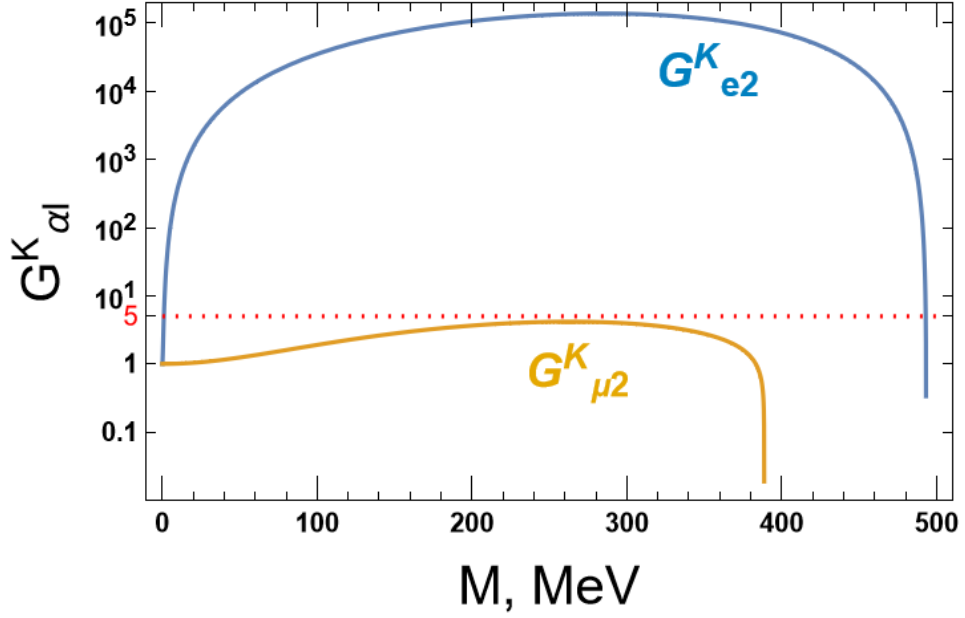}
    \caption{Dependence of the kinematic functions $G^K_{\alpha2}$, $\alpha = e,\mu$ for $K^+$-decay on the HNL mass. Note that if $M_2 = M_3 = M$, then the functions $G^K_{\alpha 2}$ and $G^K_{\alpha 3}$ coincide.}
    \label{fig:9}
\end{figure}
\begin{figure}[t]
\begin{subfigure}[t]{0.48\textwidth}
    \centering
            \includegraphics[scale=0.38]{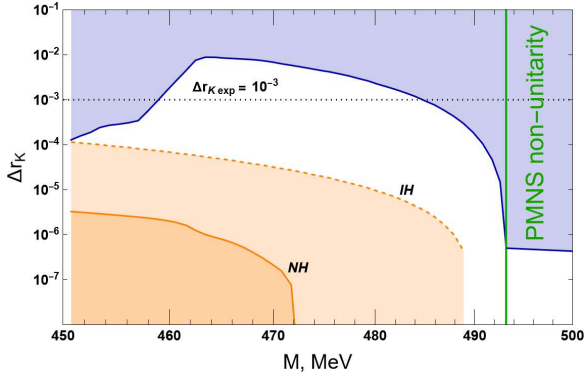}
             \caption{Lepton universality violation (LUV) parameter in $K^+$ decays with the mixing matrix corresponds to Eq.\eqref{omega22}, scenario 2. Orange lines correspond to the minimum value of LUV in the case of three HNL generations when $U_\alpha^2 = \frac{m_{1(3)}}{M_1}|U_{\alpha1(3)}|^2+|\Theta_{\alpha2}|^2+|\Theta_{\alpha3}|^2$, where $m_{1(3)} \sim 10^{-5}$ eV. Solid line denotes the case of NH, dashed line is for IH. Blue line shows a maximum value of LUV parameter for both hierarchies, see Eq.\eqref{num:r_k:kinem} for details. Horizontal dotted black line is the experimental value of $\Delta r_K \simeq 10^{-3}$, see \cite{pdg}. Vertical green line is a kinematic threshold $M=M_K-m_e$ for kaon decay $K^+ \to e^+,N_{2(3)}$.}
            \label{fig:luv_kaon}
        \end{subfigure}
    \hfill
        \begin{subfigure}[t]{0.48\textwidth}
            \centering
            \includegraphics[scale=0.38]{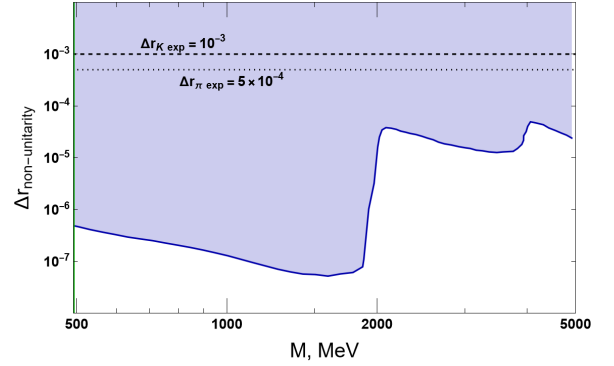}
            \caption{Lepton universality violation parameter in $\pi^+$ and $K^+$ decays when the mass of HNL is greater than all kinematic thresholds for considered decay channels. In such a case LUV occurs due to PMNS non-unitarity: $\sum_{i} |U_{\alpha i}|^2 = 1 - \sum_{I} |\Theta_{\alpha I}|^2 \neq 1$. The maximum value of $\Delta r_{non-unitary}$ is estimated from the upper experimental bound on $U^2_\mu$, see Eq.\eqref{num:r_pi} for details.}
            \label{fig:delta_r_nu}
        \end{subfigure}
    \caption{Lepton universality violation (LUV) parameters $\Delta r_\pi$ and $\Delta r_K$ as a function of the HNL masses in the case of quasidegenerate $M_2\simeq M_3 \equiv M$. The BBN bound is respected for the depicted mass interval according to the allowed domains in Fig.\ref{fig:5}. The choice of $M_1$ and the active neutrino masses for both normal and inverted hierarchies are the same as in Fig.\ref{fig:8}.}
    \label{fig:10}
    \end{figure}

\begin{table}[!h]
    \centering
        \begin{tabular}{|| c || c | c | c | c ||} 
             \hline
             $\alpha$ & $e$ & $e$ & $\mu$ & $\mu$ \\ [1ex] 
             \hline
             meson & $\pi$ & $K$ & $\pi$ & $K$ \\ [1ex]
             \hline
             $G^{meson}_{\alpha 1}-1$ & $9.6\cdot 10^{-5}$ & $9.6\cdot 10^{-5}$ & $4.8\cdot 10^{-10}$ & $2.5\cdot 10^{-9}$ \\ [1ex]
             \hline
        \end{tabular}
        \caption{Values of kinematic function $G^{M}_{\alpha 1}-1$ for $M={\pi,K}$-meson and $\alpha=e,\mu$. Mass of HNL DM fermion is $M_1=5$ keV, mass of the lightest neutrino is $10^{-5}$ eV.}
        \label{table:3}
\end{table}
As demonstrated by Fig.\ref{fig:9}, the kinematic factor $G^K_{\mu2}$ does not exceed 5, so the value of $(G^K_{\mu2}-1)\left(U_\mu^2 - |\Theta_{\mu1}|^2\right) \sim U_\mu^2 \ll 1$.
Using this approximation, the parameter of lepton universality violation (LUV) can be written in a simple form 
\begin{eqnarray}
    \Delta r_\pi &\equiv& \Delta r_{\rm non-unitary} \simeq (U_\mu^2 - |\Theta_{\mu1}|^2)-(U_e^2-|\Theta_{e1}|^2),\\
    \Delta r_K &\simeq& \Delta r_{\rm non-unitary} + G^K_{e2}\left(U_e^2 - |\Theta_{e1}|^2\right) - G^K_{\mu2}\left(U_\mu^2 - |\Theta_{\mu1}|^2\right).
\end{eqnarray}
In the following the upper and lower bounds  are denoted by $\overline{u}$ and $\underline{u}$ for $U_e^2$ and by $\overline{v}$ and $\underline{v}$ for $U_\mu^2$. Then the maximum and minimum values of the LUV parameter for the pion are
\begin{equation}
\label{num:r_pi}
    (\Delta r_\pi)_{\min} \simeq (\underline{v}-\overline{u}) + \delta_{dm}, \hspace{1cm}
    (\Delta r_\pi)_{\max} \simeq (\overline{v}-\underline{u}) + \delta_{dm} 
\end{equation}
and for the kaon ($M>430$~MeV)
\begin{eqnarray}
\label{num:r_k}
    (\Delta r_K)_{\max} &\simeq& (\overline{v}-\underline{u}) + \delta_{dm} + G^K_{e2}(\overline{u} - |\Theta_{e1}|^2) - G^K_{\mu2}(\underline{v} - |\Theta_{\mu1}|^2), \\
    (\Delta r_K)_{\min} &=& (\Delta r_K)_{\max} \text{~with replacement~} \underline{u} \leftrightarrow \overline{u},~ \overline{v} \leftrightarrow \underline{v} \nonumber
\end{eqnarray}
where 
\begin{equation}
    \delta_{dm}= |\Theta_{e1}|^2 - |\Theta_{\mu1}|^2 = \Big\{
    \begin{array}{c}
         1.12\cdot 10^{-9} ~~~\text{for NH}, \\
        -1.04\cdot10^{-9} ~~~\text{for IH}
    \end{array}  
\end{equation}
If the mass of the lightest active neutrino $m_{1(3)}=0$ (or, equivalently, if there are only two generations of HNL) then $\delta_{dm}=0$. In the considered mass range $M>410$ MeV, one can assume that $\overline{u},\overline{v} > 10^{7}$, so $\overline{u} \gg \underline{v}, \delta_{dm}$ and $\overline{v} \gg \underline{u}, \delta_{dm}$. Therefore, when the HNL mass is greater than the kinematic threshold $M_K - m_e$, it follows that $(\Delta r_{\rm non-unitary})_{\max} \simeq \overline{v}$ and $(\Delta r_{\rm non-unitary})_{\min} \simeq -\overline u$. The case $450~\text{MeV} < M \lesssim 493$ MeV is separately considered since LUV occurs due to the decay channel $K^+ \to e^+, N_{2,3}$ with a kinematic factor up to $10^5$ (see Fig.\ref{fig:9}) and in such case
\begin{equation} 
    \label{num:r_k:kinem}
    \begin{array}{ccc}
         (\Delta r_K)_{\max} & \sim & \overline{v} + G^K_{e2}(\overline{u}-|\Theta_{e1}|^2) \hspace{2cm}~\\
          (\Delta r_K)_{\min} & \sim & -(\overline{u} + G^K_{\mu2}\overline{v})  + G^K_{e2}(\underline{u}-|\Theta_{e1}|^2)
    \end{array}
     ~\text{for}~~450~\text{MeV} < M < 493~\text{MeV}.
\end{equation}
These boundaries are shown in Fig.\ref{fig:luv_kaon}. Maximum value of LUV parameter for HNL mass greater than the kinematic threshold ($M>493$ MeV) is shown in Fig.\ref{fig:delta_r_nu}.
\section{Summary}

The most general case of extending the lepton sector of the SM by three sterile Majorana neutrinos in order to generate the masses of standard neutrinos using the see-saw mechanism allows one to build a structured hierarchy of mixing parameters within a well-defined basis for mass states. Cosmological limitations on the lifetime and energy fraction of the lightest mass state of neutral heavy leptons, considered as a dark matter particle, restrict the mass to vary in an interval of 0.4-40 keV within the sensitivity of modern experiments, which allows the use of preferred forms of active and heavy neutrino mixing matrices for the analysis of data from experiments with extracted beams and colliders.

Limitations of the possible form of the mixing matrix in the see-saw type I models, which are of undoubted interest, can be obtained by combining constraints on the matrix elements imposed from above by the absence of signals on colliders and beam dump experiments, while constraints on the mixing from below are provided by baryogenesis scenarios within the framework of the Big Bang concept, as well as restrictions on flavor oscillations. However, the numerical values for the boundaries strongly depend on the mixing scenario within a particular model which, from a technical point of view, is ambiguously implemented by a certain choice of the $\Omega$-matrix in the $M_D - U_{PMNS} - U_N$ connection, Eq.\eqref{eq:md-omega-un}. The lepton universality violation parameter in $K^\pm$ and $\pi^\pm$ decays is sensitive 
to the neutral heavy leptons due to their additional contributions to meson decay widths, dependent on the mixing matrix.

Two mixing scenarios analysed above are rather different, in the "fine-tuned" scenario 1 the mixing matrix depends on the mass ratio $(m_i/M_i)^{1/2}$ of active neutrino and HNL, which suppresses any HNL production or decay channel and imposes rather strict restrictions on $N_1$ from the cosmological lifetime and DM energy fraction requirements, not affecting $N_2$ and $N_3$ masses.  In the scenario 2 with three additional parameters for the $N_2$ and $N_3$ mixing, intensively discussed in the literature, a sort of tuning is needed to generate small masses of active neutrinos by means of see-saw type I and the baryon asymmetry of the Universe by means of $N_2-N_3$ flavor oscillation mechanism. HNL production is enhanced by the mixing parameter $e^{{\rm Im}(\omega)}$ and stronger restrictions on the $N_2$ and $N_3$ masses ($N_2\sim N_3$) from below are imposed not affecting to large extent $N_1$ which has a negligible mixing.
Strong hierarchy of $\Theta_{eI}$ and $\Theta_{\mu I}$, $\Theta_{\tau I}$ elements of the mixing matrix naturally appears in scenario 1 and can be easily configured for the NH case in scenario 2. The cosmological upper bound on the HNL lifetime 
is critical for determining the bound on the masses of $N_2$ and $N_3$, so their lifetime was carefully evaluated taking into account the three-particle leptonic decay channels which are dominant below the thresholds of the two-particle channels. Our main results can be summarized as
\begin{itemize}
    \item In the range of HNL masses $120~\text{MeV} < M < 140~\text{MeV}$ (close to the mass of $\pi$-meson) a small window for the $U^2_e$ consistent with the data on the missing energy reconstruction and search for displaced vertices  was found in the case of a normal hierarchy (NH): $3\cdot 10^{-8} < U_e^2 < 2\cdot10^{-7}$.
    \item There is a possible range of parameters consistent with the experimental upper bound for $U_e^2$, $U_\mu^2$ and the limitations of Big Bang Nucleosynthesis (BBN) with the following dependencies on the mass hierarchy:
    \begin{eqnarray*}
        M &>& 430~\text{MeV} ~~~ \text{for $U_e^2$ with NH}, ~~~ M > 430~\text{MeV} ~~~\text{for $U_e^2$ with IH}; 
        \\
        M &>& 290~\text{MeV} ~~~ \text{for $U_\mu^2$ with NH}, ~~~ M > 300~\text{MeV} ~~~\text{for $U_\mu^2$ with IH}.
    \end{eqnarray*}
    Combining this constraints, we can conclude that $M>430$ MeV in addition to the allowed window of HNL mass mentioned above.
    \item In the model with three generations of sterile Majorana neutrinos where the lightest active neutrino mass is $m_{1} \sim 10^{-5}$ eV  and $M_{DM} \sim 1$ keV the lower limit for HNL mass coming from the $U_e^2$ BBN bound $M > 430$ MeV (NH) is raised in comparison with the simplified model with two generations of HNL where $m_{1} = 0$ and $M > 170$ MeV.

\item
The lepton universality violation parameter $\Delta r_M$, $M=\pi^\pm, K^\pm$ does not exceed ${\cal O}(10^{-4})$ at $M_2$ and $M_3$ masses of the order of 10$^2$ MeV in the scenario 1, demonstrating very high sensitivity to the BBN lifetime restrictions ($M>480$ MeV for IH and $M>830$ MeV for NH). In allowed BBN region LUV parameter does not exceed $\mathcal{O}(10^{-10})$ for NH and $\mathcal{O}(10^{-7})$ for IH. The situation is more involved in the three-parametric scenario 2 where the lifetime bound is given by an exclusion contours on the mixing -- mass, ${\rm Im}(\omega)-M$ plane.

Using the theoretical SM value for $R_\pi$ and $R_K$ from \cite{ktheor,pitheor} and the experimental value from Particle Data Group \cite{pdg}, the lepton universality violation parameter for $\pi^\pm$ is found to be $\Delta r_\pi=(-4\pm 3)\times 10^{-4}$ and $\Delta r_K=(4\pm 4)\times 10^{-3}$ for $K^\pm$.
A maximal value of the lepton universality violation parameter $\Delta r_K$ is comparable or exceeds the experimental value in the HNL mass range between 460 -- 485~MeV. It follows that the currently achieved level of experimental accuracy is quite moderate, as a result of which future experiments of high precision have a great potential for the discovery of BSM physics lepton mixing scenarios.
\end{itemize}
Some differences between the BBN exclusion contours obtained in the literature and in evaluations above can take place due to approximations for the explicit form of mixing matrices for three HNL generations in the scenarios under consideration, and calculations of the widths of three-particle HNL decays beyond the Dirac limit. 
The HNL lifetime bound is essential for manipulations with displaced vertices. If a lifetime bound is denoted by $\tau_0$ and a decay width is factorized in the form of a mixing factor times matrix element squared, $\Gamma_0=|U_\alpha|^2 \, M^2$, then a bound for the mixing factor is $|U_\alpha|^2> 1/(\tau_0 M^2)$. This qualitative estimate gives a shift of the $|U_\alpha|^2$ contour upwards (downwards) when either the lifetime bound or the matrix element squared decreases (increases).
For HNL masses less than the threshold of around 0.14 GeV, see \cite{improved-bbn}, the BBN contour in Fig.\ref{fig:5} can be shifted upwards, which is explained by a step-like approximation of the upper limit for the lifetime $\tau_0$ increase to approximately 0.1 sec. At masses exceeding the threshold of 0.14 GeV the contour in Fig.\ref{fig:5} can be shifted downwards, which is explained by different ways of taking into account the permissible areas for ${\rm Im} \, \omega$ and $M$, see Fig.\ref{fig:7}. When generating a contour in this paper, the lower permissible boundary on the ${\rm Im} \, \omega - M$ plane exactly corresponds to a displacement along the curve in Fig.\ref{fig:7}, whereas for the case of two HNL generations either asymptotic behavior is used, see Eq.(\ref{asymptotic}), or the limiting case of the so-called "dominant mixing" is taken, when the ratios $|U_e|^2 \, : \, |U_\mu|^2 : \, |U_\tau|^2$  have the form of a ratio of some numerical constants, which is equivalent to a step-function approximation. Departures in the case of inverse neutrino mass hierarchy are a consequence of completely different structure of mixing matrices for NH and IH in our case. 
Some displacement of the contours appears also due to contributions to the width by the interference terms or different signs of these terms in the three-particle decays beyond the Dirac limit, however, against the background of two-particle contributions above the thresholds, this displacement is not very significant. As a result of evaluations, in our case the open windows for HNL masses are slightly smaller  and for larger values of $M_{\rm HNL}$, the lower permissible limit is changed. 
Significant changes occur with the three-generation seesaw bound curves compared to the two-generation absolute seesaw bound curves due to an additional term  in the parameters $U^2_e$ and $U^2_\mu$ which includes non-zero active neutrino masses.
The abovementioned changes do not affect the mass range, which is most interesting for observing the violation of lepton universality.

\vskip 3mm
{\it Acknowlegement \hskip 2mm}
The work of M.D. was supported by the Russian Science Foundation Grant No. 22-12-00152.

\section*{Appendix A. Amplitudes for the three-particle leptonic decays}
Models with Majorana fermions have a number of features of the use of diagram technique, see \cite{denner, haber}. Diagram technique for Majorana fermions proposed in \cite{haber} where the charge conjugation matrix was explicitly included in the Feynman rules, has been modified in \cite{denner} where standard propagators are used, vertices do not include the charge conjugation matrix and a specific {\it fermion flow} defined for fermion lines is introduced. This Appendix provides the specifics of calculations in the cases under consideration.
\begin{table}[h]
    \centering
    \begin{tabular}{|m{10cm}|m{4cm}|}
        \hline
        Trace/combination of traces & Result of calculation \\
        \hline
        $Tr[\hat{A}\gamma_\alpha P_{L(R)} \hat{B}\gamma^\beta P_{L(R)} \hat{C} \gamma^\alpha P_{L(R)} \hat{D} \gamma_\beta P_{L(R)}~]$ & $-16 (A C)(B D)$ \\
        \hline 
        $Tr[\hat{A}\gamma_\alpha P_{L(R)} \hat{B} \gamma_\beta P_{L(R)} ] \cdot Tr[\hat{C}\gamma^\alpha P_{L(R)} \hat{D} \gamma^\beta P_{L(R)}]$ & $16(A C) (B D)$ \\
        \hline 
        $Tr[\hat{A}\gamma_\alpha P_{L} \hat{B} \gamma_\beta P_{L} ] Tr[\hat{C}\gamma^\alpha P_{R} \hat{D} \gamma^\beta P_{R}]$ & $16 (A D) (B C)$\\
        \hline 
        $Tr[\hat{A}\gamma_\alpha P_{R(L)} \hat{B} P_{L(R)} \hat{C} \gamma^\alpha P_{L(R)} \hat{D} P_{R(L)}]$ & $8 (A D) (B C)$ \\
        \hline 
        $Tr[\hat{A}\gamma_\alpha P_{L(R)} \hat{B} \gamma_\beta P_{L(R)} ]~Tr[\gamma^\alpha P_{L(R)} \gamma^\beta P_{R(L)}]$ & $-8 (A B)$\\
        \hline 
        $Tr[\hat{A}\gamma_\alpha P_{L(R)} \hat{B} \gamma_\beta P_{L(R)} ]~Tr[\gamma^\alpha P_{R(L)} \gamma^\beta P_{L(R)}]$ & $-8 (A B)$ \\
        \hline 
        $Tr[\hat{A}\gamma_\alpha P_{L(R)} \gamma_\beta P_{R(L)} \gamma^\alpha P_{L(R)} \hat{D} \gamma^\beta P_{L(R)}]$ & $8 (AD)$\\
        \hline 
    \end{tabular}
    \caption{Formulas for all nonzero traces (combination of traces) of gamma matrices arising during the calculation in cases 1,2,3}.
    \label{table:2}
\end{table}
\begin{table}
    \centering
        \begin{tabular}{ c  c }
        \vspace{1cm}
            \parbox{2cm}{\begin{fmfgraph*}(70,70)
                    \fmfleft{i1}                    
                    \fmfright{o1,o2} \fmf{photon,label.side=left,label=$Z$}{i1,w1} \fmf{plain,label.side=right,label=$\upnu_j$}{w1,o1}                    \fmf{plain,label.side=left,label=$N_I$}{w1,o2}
                    \fmfv{lab=$\circlearrowright$}{w1}
            \end{fmfgraph*}} & $~~~-i\frac{g}{2 c_W}\gamma^\mu \left[ (U^\dagger\Theta)^*_{jI} P_L - (U^\dagger\Theta)_{jI} P_R \right]$ \\
        \vspace{1cm}
            \parbox{2cm}{\begin{fmfgraph*}(70,70)
                \fmfleft{i1}
                \fmfright{o1,o2}
                \fmf{photon,label.side=right,label=$Z$}{i1,w1}
                \fmf{plain,label.side=right,label=$\upnu_i$}{w1,o1}
                \fmf{plain,label.side=left,label=$\upnu_i$}{w1,o2}
                \fmfv{lab=$\circlearrowright$}{w1}
            \end{fmfgraph*}} & $-i\frac{g}{2 c_W}\gamma^\mu (P_L - P_R)$ \\
        \vspace{1cm}
             \parbox{2cm}{\begin{fmfgraph*}(70,70)
                \fmfleft{i1}
                \fmfright{o1,o2}
                \fmf{photon,label.side=left,label=$W^{+}$}{i1,w1}
                \fmf{plain,label.side=right,label=$N_I$}{w1,o1}
                \fmf{fermion,label.side=left,label=$l_\alpha^{-}$}{w1,o2}
                \fmfv{lab=$\circlearrowright$}{w1}
            \end{fmfgraph*}}  & $-i\frac{g}{\sqrt{2}} \Theta_{\alpha I}\gamma^\mu P_L$ \\
        \vspace{1cm}
             \parbox{2cm}{\begin{fmfgraph*}(70,70)
                \fmfleft{i1}
                \fmfright{o1,o2}
                \fmf{photon,label.side=left,label=$W^{-}$}{i1,w1}
                \fmf{plain,label.side=right,label=$N_I$}{w1,o1}
                \fmf{fermion,label.side=right,label=$l_\alpha^{+}$}{o2,w1}
                \fmfv{lab=$\circlearrowright$}{w1}
            \end{fmfgraph*}}  & $+i\frac{g}{\sqrt{2}} \Theta_{\alpha I}^*\gamma^\mu P_R$ \\
        \vspace{1cm}
             \parbox{2cm}{\begin{fmfgraph*}(70,70)
                \fmfleft{i1}
                \fmfright{o1,o2}
                \fmf{photon,label.side=left,label=$Z$}{i1,w1}
                \fmf{fermion,label.side=left,label=$l_\alpha^+$}{o1,w1}
                \fmf{fermion,label.side=left,label=$l_\alpha^-$}{w1,o2}
                \fmfv{lab=$\circlearrowright$}{w1}
            \end{fmfgraph*}}  & $~~-i\frac{g}{2 c_W}\gamma^\mu \left((2s_W^2-1)P_L + 2s_W^2 P_R\right) $
        \\
        \end{tabular}
        \caption{Feynman rules for Majorana fermions used in calculations of pure leptonic decay amplitudes. The direction of the fermion flow is chosen from the bottom up, as indicated by an arrow near the vertex, $P_{R,L}=(1\pm\gamma_5)/2$, $s_W=\sin \theta_W$. Nonstandard pseudovector structure of $\nu \bar \nu Z$ vertex, for example, follows from the equality of vector current to zero for Majorana fermion.}
        \label{table:1}
    \end{table}
    
\textbf{Case 1}:
Here we choose the direction of the \textit{fermion flow} (see \cite{denner}) as shown on diagram A1, then
\begin{figure}[h!]
    \begin{center}
       \begin{fmfgraph*}(120,120)
            \fmfstraight
            \fmfleft{i1,i2,id1,id2,i3,i4,i5}
            \fmfright{o1,o2,od1,od2,o3,o4,o5}
            \fmf{plain,tension=1,label=$N_I(p)$, label.side=right, label.dist=0.1cm}{v2,i4}
            \fmf{plain,label=$\upnu_i(k_1)$, label.side=right, label.dist=0.1cm}{o4,v2}
            \fmffreeze
            \fmf{plain}{o1,v1,o3}
            \fmf{plain,tension=3,label=$~\upnu_j(k_3)$, label.side=right, label.dist=0.1cm}{o1,v1}
            \fmf{plain,label=$~\upnu_j(k_2)$, label.side=right, label.dist=0.1cm}{v1,o3}
            \fmf{photon, tension=2,label=$Z$, label.side=right, label.dist=0.2cm}{v2,v1}
            \fmfv{lab=$\leftarrow$}{v2}
            \fmfv{lab=$\circlearrowleft$}{v1}
        \end{fmfgraph*}
    \vspace{0.2cm}\\
    Diagram A1. Diagram for the case 1 with with an explicitly specified choice of fermion flow direction indicated by arrows.
    \end{center}
\end{figure}

\begin{eqnarray}
    M_{3\upnu} = -i\sqrt{2}G_F (\overline{v}_N \gamma_\mu \left[(U^\dagger\Theta)^*_{iI} P_L - (U^\dagger\Theta)_{iI} P_R \right] v_1)(\overline{u}_3 \gamma^\mu (P_L-P_R) v_2)
\end{eqnarray}
Denoting the terms of the full decay amplitude as
\begin{eqnarray*}
    M_{LL} &=& -i\sqrt{2} G_F (U^\dagger\Theta)_{iI}^*(\overline{v}_N \gamma_\mu P_L v_1) (\overline{u}_3 \gamma^\mu P_L v_2) \\
    M_{LR} &=& -i\sqrt{2} G_F (U^\dagger\Theta)_{iI}^*(\overline{v}_N \gamma_\mu P_L v_1) (\overline{u}_3 \gamma^\mu P_L v_2) \\
    M_{RL} &=& -i\sqrt{2} G_F (U^\dagger\Theta)_{iI}^*(\overline{v}_N \gamma_\mu P_L v_1) (\overline{u}_3 \gamma^\mu P_L v_2) \\
    M_{RR} &=& -i\sqrt{2} G_F (U^\dagger\Theta)_{iI}^*(\overline{v}_N \gamma_\mu P_L v_1) (\overline{u}_3 \gamma^\mu P_L v_2), 
\end{eqnarray*}
one can observe that there is no interference in the approximation of zero masses of active neutrinos in the final state
\begin{eqnarray*}
    &&Tr[(\hat{p}-M_I)\gamma_\mu P_{L(R)} \hat{k}_1\gamma_\nu P_{R(L)}] = 0,\\
    &&Tr[\hat{k}_3\gamma^\mu P_{L(R)} \hat{k}_2 \gamma^\nu P_{R(L)}] = 0.
\end{eqnarray*}

For the amplitude terms we get 
\begin{eqnarray*}
    |M_{LL}|^2 &=& 2 G_F^2 |(U^\dagger\Theta)_{iI}|^2 Tr[(\hat{p}-M_I)\gamma_\mu P_{L} \hat{k}_1\gamma_\nu P_{L}]~Tr[\hat{k}_3\gamma^\mu P_{L} \hat{k}_2 \gamma^\nu P_{L}]=\\
    &=& 32 G_F^2 |(U^\dagger\Theta)_{iI}|^2(p k_3)(k_1 k_2),\\
    |M_{RL}|^2 &=& 2G_F^2 |(U^\dagger\Theta)_{iI}|^2 Tr[(\hat{p}-M_I)\gamma_\mu P_{R} \hat{k}_1\gamma_\nu P_{R}]~Tr[\hat{k}_3\gamma^\mu P_{L} \hat{k}_2 \gamma^\nu P_{L}]=\\ 
    &=& 32 G_F^2 |(U^\dagger\Theta)_{iI}|^2(p k_2) (k_1 k_3), \\
    |M_{LR}|^2 &=& 2 G_F^2 |(U^\dagger\Theta)_{iI}|^2 Tr[(\hat{p}-M_I)\gamma_\mu P_{L} \hat{k}_1\gamma_\nu P_{L}]~Tr[\hat{k}_3\gamma^\mu P_{R} \hat{k}_2 \gamma^\nu P_{R}]=\\
    &=& 32 G_F^2 |(U^\dagger\Theta)_{iI}|^2(p k_2) (k_1 k_3), \\
    |M_{RR}|^2 &=& 2 G_F^2 |(U^\dagger\Theta)_{iI}|^2 Tr[(\hat{p}-M_I)\gamma_\mu P_{R} \hat{k}_1\gamma_\nu P_{R}]~Tr[\hat{k}_3\gamma^\mu P_{R} \hat{k}_2 \gamma^\nu P_{R}]=\\
    &=& 32 G_F^2 |(U^\dagger\Theta)_{iI}|^2 (p k_3)(k_1 k_2),\\
\end{eqnarray*}
and finally for the full squared amplitude 
\begin{equation*}
    |M_{3\upnu}|^2 = 64 G_F^2 |(U^\dagger\Theta)_{iI}|^2~\left[ (p k_3) (k_1 k_2) + (p k_2) (k_1 k_3) \right].
\end{equation*}
Identical particles in the final state give an additional factor $\frac{1}{2}$ for decay width. Note that for Dirac active neutrino terms $LR$ and $RR$ do not arise and the factor in front of right-hand side will be 32. It is also necessary to take into account the charge conjugate mode by multiplying by two. There is no charge conjugated mode for Majorana neutrinos. In the Dirac limit, see \cite{Gorbunov-Shaposh}, Eq.(3.5) for $\Gamma(N\to \nu\nu\bar{\nu})$, summation over flavor indices and multiplication by a factor of 2 for charge conjugated final states gives the same result as Eq.\eqref{decay:3nu} above.

\textbf{Case 2.} Acting by the same rules in combination with Fiertz transformations, see Table 2, the amplitudes $M_1$ and $M_2$ for the diagram with intermediate $W^+$ and the diagram with intermediate $W^-$ can be written as
\begin{figure}[h!]
    \begin{center}
        \begin{fmfgraph*}(120,120)
            \fmfstraight
            \fmfleft{i1,i2,id1,id2,i3,i4,i5}
            \fmfright{o1,o2,od1,od2,o3,o4,o5}
            \fmf{plain,tension=1.5,label=$N_I(p)$, label.side=right, label.dist=0.1cm}{v2,i4}
            \fmf{fermion,label=$l_\alpha^+(p_1)$, label.side=right, label.dist=0.1cm}{o4,v2}
            \fmffreeze
            \fmf{plain}{o1,v1,o3}
            \fmf{plain,tension=3,label=$\upnu_k(k)$, label.side=right, label.dist=0.1cm}{o1,v1}
            \fmf{fermion,label=$l_\beta^-(p_2)$, label.side=right, label.dist=0.1cm}{v1,o3}
            \fmf{photon, tension=2,label=$W^+$, label.side=right, label.dist=0.2cm}{v2,v1}
            \fmfv{lab=$\circlearrowleft$}{v1}
            \fmfv{lab=$\leftarrow$}{v2}
        \end{fmfgraph*}
            \hspace{3cm}
        \begin{fmfgraph*}(120,120)
            \fmfstraight
            \fmfleft{i1,i2,id1,id2,i3,i4,i5}
            \fmfright{o1,o2,od1,od2,o3,o4,o5}
            \fmf{plain,tension=1.5,label=$N_I(p)$, label.side=right, label.dist=0.1cm}{v2,i4}
            \fmf{fermion,label=$l_\beta^-(p_2)$, label.side=left, label.dist=0.1cm}{v2,o4}
            \fmffreeze
            \fmf{plain}{o1,v1,o3}
            \fmf{plain,tension=3,label=$\upnu_k(k)$, label.side=right, label.dist=0.1cm}{o1,v1}
            \fmf{fermion,label=$l_\alpha^+(p_1)$, label.side=left, label.dist=0.1cm}{o3,v1}
            \fmf{photon, tension=2,label=$W^-$, label.side=right, label.dist=0.2cm}{v2,v1}
            \fmfv{lab=$\circlearrowleft$}{v1}
            \fmfv{lab=$\leftarrow$}{v2}
        \end{fmfgraph*}
    \end{center}
    \vspace{0.2cm}
    \begin{center}
        Diagrams A2. Diagrams for case 2 with an explicitly specified choice of fermion flow direction (thin arrows).
    \end{center}
\end{figure}
\begin{eqnarray*}
    M_1 &=& i2\sqrt{2} G_F \Theta^*_{\alpha I} U_{\beta i} (\overline{v}_N \gamma_\mu P_L v_1) (\overline{u}_\nu \gamma^\mu P_R v_2), \\
    M_2 &=& i2\sqrt{2} G_F \Theta_{\beta I} U^*_{\alpha i} (\overline{v}_N \gamma_\mu P_R v_2) (\overline{u}_\nu \gamma^\mu P_L v_1),
\end{eqnarray*}
the squared terms are
\begin{eqnarray*}
    |M_1|^2 &=& 128 G_F^2 |\Theta^*_{\alpha I}|^2 |U_{\beta i}|^2 (p p_2)(p_1 k), \\
    |M_2|^2 &=& 128 G_F^2 |\Theta_{\beta i}|^2 |U_{\beta i}|^2 (p p_1) (p_2 k), \\ 
    M_1 M_2^\dagger &\sim& Tr[(\hat{p}-M_I)\gamma_\mu P_L (\hat{p}_1 - m_1) \gamma^\nu P_L \hat{k}\gamma^\nu P_R (\hat{p}_2 - m_2) \gamma_\nu P_R]=0,
\end{eqnarray*}
and the squared amplitude $|M_W|^2=|M_1 + M_2|^2$ for the case 2 has the form 
\begin{equation*}
    |M_W|^2 = 128 G_F^2 \left[|\Theta_{\alpha I}|^2 |U_{\beta i}|^2 (p p_2)(p_1 k) + |\Theta_{\beta I}|^2 |U_{\alpha i}|^2 (p p_1)(p_2 k)\right].
\end{equation*}
The decay width for all lepton masses nonzero is given by Eq.\eqref{eq:width:case3}.\\

\textbf{Case 3}.
For the three diagrams in this case, the notation $M_{3_{XY}}$, $XY={LL,LR,RL,RR}$ is used for the neutral current diagram and the notation $M_1,M_2$ is used for $W^+$ and $W^-$ exchange diagrams, respectively. Then the full amplitude $|M_{\rm WZ}|$ contains 21 terms: 6 terms for the squared diagrams and $\frac{36-6}{2}=15$ terms for the interferences. Moreover, we need to change the fermion flow depending on each specific interference term to build a trace of gamma matrices. For instance, we need to change the fermion flow for $Z \upnu l$-vertex in the diagram with $W^+$ in order to calculate $M_1 M_{3_{XY}}^\dagger$. Full interference term looks as 
\begin{equation*}
    M_{\rm interf}^2 \equiv \sum \limits_{XY} 2Re\left[M_1 {({M_3}^\prime_{XY})}^\dagger\right] + \sum \limits_{XY} 2Re\left[M_2^\prime {({M_3}^{\prime \prime}_{XY})}^\dagger \right].
\end{equation*}
Symbols "$\prime$" and "$\prime \prime $" mark those diagrams in which it was necessary to change the fermion flow for calculation,
\begin{eqnarray*}
    |M_{\rm WZ}|^2 = |M_Z|^2 + |M_W|_{\alpha=\beta}^2 + M_{\rm interf}^2,
\end{eqnarray*}
where 
\begin{eqnarray*}
    |M_Z|^2 &=& 128 G_F^2 \left(\mathcal{C}_1^2 + \mathcal{C}_2^2\right)~|(U\Theta)_{iI}|^2 ~\left[(p p_2)(p_1 k)+(p p_1)(p_2 k)\right] + \\
    && + 128G_F^2 (4 \mathcal{C}_1\mathcal{C}_2)~|(U^\dagger\Theta)_{iI}|^2 ~\frac{m_\alpha^2}{2} (pk),\\
    |M_W|_{\alpha=\beta}^2 &=& 128 G_F^2~|\Theta_{\alpha I}|^2 |U_{\alpha i}|^2 \cdot \left[(p p_2)(p_1 k) + (p p_1)(p_2 k)\right], \\
    - M_{\rm interf}^2 &=& 128 G_F^2 \mathcal{C}_1 \cdot 2\left[(U^\dagger \Theta)_{iI} \Theta_{\alpha I}^* U_{\alpha i}~(p p_2) (p_1 k) + (U^\dagger \Theta)^*_{iI} \Theta_{\alpha I} U_{\alpha i}^*~(p p_1) (p_2 k) \right] + \\
    && + 128 G_F^2 \mathcal{C}_2 \cdot 2\left[(U^\dagger \Theta)_{iI} \Theta_{\alpha I}^* U_{\alpha i} + (U^\dagger \Theta)^*_{iI} \Theta_{\alpha I} U_{\alpha i}^*\right] \cdot \frac{m_\alpha^2}{2} (pk), 
\end{eqnarray*}
and we use notation ($s_W=\sin \theta_W$ where $\theta_W$ is the Weinberg angle)
\begin{eqnarray*}
    \mathcal{C}_1=\left(s_W^2-\dfrac{1}{2}\right), ~~~ \mathcal{C}_2=s_W^2.
\end{eqnarray*}
The square of the full amplitude is simplified after summing by the neutrino mass states, Eq.(25),
\begin{eqnarray}
    \sum \limits_i |M_{\rm WZ}|^2 &=& 128 G_F^2 \left[\left(\mathcal{C}_1^2 + \mathcal{C}_2^2\right) \sum \limits_\beta |\Theta_{\beta I}|^2 + \left(1 - 2\mathcal{C}_1\right) |\Theta_{\alpha I}|^2 \right] ~\left[(p p_2)(p_1 k)+(p p_1)(p_2 k)\right] + \\
    && + 128 G_F^2 \left[\left(4\mathcal{C}_1\mathcal{C}_2\right)\sum \limits_\beta |\Theta_{\beta I}|^2 - 4\mathcal{C}_2 |\Theta_{\alpha I}|^2\right] ~ \frac{m_\alpha^2}{2} (p k) ,
\end{eqnarray}
the decay width for lepton nonzero masses is given by Eq.\eqref{m2wz}. Note that for Dirac neutrinos, opposite signs of interference terms above appear.

\section*{Appendix B. Kinematics of $1\to 3$ decay with mass terms}
This Appendix contains the details of integration by the Dalitz plot in the general case of non-zero masses using invariant variables. For the process $N_I(p) \to a(p_1),b(p_2),c(p_3)$ invariant kinematic variables are defined as $S_1 = (p-p_3)^2 = (p_1+p_2)^2$, $S_2 = (p-p_1)^2 = (p_2 + p_3)^2$,
    $S_3 = (p-p_2)^2 = (p_3 + p_1)^2$ which are then redefined to dimensionless variables
for convenience, $s_i \equiv \frac{S_i}{M_I^2}$, $r_i \equiv \frac{m_i^2}{M_I^2}$ where $m_i^2 = p_i^2$, $i=1,2,3$ and the width
\begin{eqnarray*}
    \Gamma(N_I \to a,b,c) &=& \frac{1}{256 \pi^3 M_I^3} \int dS_1 dS_2 |M(S_1,S_2)|^2 ~\Uptheta\left[-G(S_1,S_2, M_I^2, m_2^2,m_1^2,m_3^2)\right]=\\
    &=& \frac{M_I^5}{256 \pi^3}\int  ds_1 ds_2 |M(s_1,s_2)|^2 ~\Uptheta\left[-G(s_1,s_2,1,r_2,r_1,r_3)\right]
\end{eqnarray*}
where $\Uptheta[x]$ is the step function and $G$ is the Gram determinant
\begin{eqnarray*}
    G(x,y,z,a,b,c) &=& x^2 y+x y^2-x y (z+a+b+c)-\\
    &&-b c (x+y+z+a)-z a (x+y+b+c)+\\
    &&+ x z c+x a b+y z c+y a c+z^2 a+z a^2+b^2 c+b c^2.
\end{eqnarray*}
The equations for the boundaries of the physical region in terms of invariants are obtained requiring $G(s_1,s_2,1,r_2,r_1,r_3) = 0$, so for $s_1$ they are
\begin{eqnarray*}
    s_1^\pm = 1 + r_3 - \frac{1}{2 s_2} \left((1-r_1 + s_2)(s_2 - r_2 + r_3) \mp \lambda^{1/2}(s_2,1,r_1)\lambda^{1/2}(s_2,r_2,r_3)\right)
\end{eqnarray*}
and for $s_2$ they are given by
\begin{equation*}
    (m_2 + m_3)^2 \leq S_2 \leq (M_I - m_1)^2 ~~~ \text{or} ~~~ (\sqrt{r_2}+\sqrt{r_3})^2 \leq s_2 \leq (1-\sqrt{r_1})^2
\end{equation*}
so the width
\begin{equation*}
     \Gamma(N_I \to a,b,c) = \frac{M_I^5}{256 \pi^3}\int \limits_{(\sqrt{r_2}+\sqrt{r_3})^2}^{(1-\sqrt{r_1})^2} ds_2 \int \limits_{s_1^{-}}^{s_1^{+}} ds_1 |M(s_1,s_2)|^2 
\end{equation*}
\end{fmffile}

\end{document}